\documentclass[journal,twoside]{asmb}

\begin{document}
\title{On Effective Secrecy Throughput of Underlay Spectrum Sharing $\alpha-\mu$/ M\'alaga Hybrid Model under Interference-and-Transmit Power Constraints}

\author[1]{Md. Ibrahim}
\author[2]{A. S. M. Badrudduza}
\author[3]{Md. Shakhawat Hossen}
\author[4]{M. K. Kundu}
\author[5]{Imran Shafique Ansari}

\affil[1]{Department of Electrical \& Electronic Engineering, Rajshahi University of Engineering \& Technology (RUET), Rajshahi-6204, Bangladesh}
\affil[2,3]{Department of Electronics \& Telecommunication Engineering, RUET}
\affil[4]{Department of Electrical \& Computer Engineering, RUET}
\affil[5]{James Watt School of Engineering, University of Glasgow, Glasgow G12 8QQ, United Kingdom}

\twocolumn[
\begin{@twocolumnfalse}
\maketitle
\begin{abstract}

\textbf{Abstract:} The underlay cognitive radio-based hybrid radio frequency / free-space optical (RF / FSO) systems have been emerged as a promising technology due to its ability to eliminate spectrum scarcity and spectrum under-utilization problems. Consequently, this work analyzes the physical layer security aspects of a cognitive RF / FSO hybrid network that includes a primary user, a secondary source, a secondary receiver, and an eavesdropper where the secret communication takes place between two legitimate secondary peers over the RF and FSO links simultaneously, and the eavesdropper can overhear the RF link only. In particular, the maximum transmit power limitation at the secondary user as well as the permissible interference power restriction at the primary user are also taken into consideration. All the RF links are modeled with $\alpha$-$\mu$ fading whereas the FSO link undergoes M\'alaga ($\mathcal{M}$) turbulence with link blockage and pointing error impairments. At the receiver, the selection combining diversity technique is utilized to select the signal with the best electrical signal-to-ratio (SNR). Moreover, the closed-form expressions for the secrecy outage probability, probability of strictly positive secrecy capacity, and effective secrecy throughput are derived to analyze the secrecy performance. Besides, the impacts of fading, primary-secondary interference, detection techniques, link blockage probability, atmospheric turbulence, and pointing error are examined. Finally, Monte-Carlo simulations are performed to corroborate the derived expressions.
\end{abstract}

\end{@twocolumnfalse}
]
\section{Introduction}

\subsection{Background and Related Works}

With the rapid increase in wireless devices, the shortage of spectrum owing to the data traffic reflects a significant threat in contemporary technology \cite{haykin2005cognitive,9508176}. In order to meet the requirements, several optimistic technologies have been developed recently. Among these, both cognitive radio network (CRN) and free-space optical (FSO) systems have received considerable attention from the research communities \cite{6952039,erdogan2019error}. Although FSO communication appears with specific merits, such as high bandwidth, inherent security, and cost-effective operation, several issues like atmospheric turbulence and pointing error can affect the secure transmission greatly \cite{juel2021secrecy}. Furthermore, the probability of transmission blockage due to extreme line-of-site (LOS) necessity in FSO link leads to the system failure \cite{djordjevic2016outage}. Hence, utilizing radio frequency (RF) technology as an alternative option along with the FSO system assures successful transmission since the FSO link delivers high data speed while the RF link is independent of the atmospheric turbulence and weather conditions \cite{ansari2013performance}.

Due to the increased demand for faster data rates in wireless communication, hybrid networks, which consist of RF and FSO systems, are deployed utilizing a variety of techniques. An adaptive coding method in order to enhance the system performance of RF / FSO hybrid network was considered in \cite{makki2016performance,khan2017adaptive}. Nonetheless, the performance of these systems is highly dependent on the existence of feedback information at the receiver \cite{khan2017adaptive}. Another attractive feature is the switching mechanism that uses RF link when FSO link is down \cite{usman2014practical}. Since one link is underutilized at a time, there is a significant amount of bandwidth wasted \cite{kaushal2016optical}. To extend the communication range, several existing works like \cite{bag2018performance,yang2017unified,singya2020performance,7881143} utilize relaying protocols in mixed RF-FSO models. However, weather conditions (i.e. fog) can influence the performance of dual-hop networks badly. Unlike \cite{usman2014practical,kaushal2016optical}, the considered hybrid RF / FSO model dispatches identical information via two separate links simultaneously and combines the received signal using the diversity combining method. Again, the mixed RF-FSO system differs completely from the hybrid RF / FSO model in terms of system configuration. Recently, a comprehensive analysis on hybrid RF / FSO model was performed in \cite{odeyemi2019selection,shakir2019performance} where the hybrid system exhibits better link availability compared to the individual links.

In CRN, an unlicensed user gets the permission of accessing the same spectrum of the licensed user through different spectrum sharing methods \cite{varshney2017cognitive}. More particularly, in the underlay approach, the secondary user ($SU$) makes use of the primary user's ($PU$) shared spectrum as long as the interference occurring at SU is less than the permitted threshold value \cite{erdogan2019error}. Considering CR technology in the RF links, the system performance of mixed RF-FSO systems are investigated in \cite{cvetkovic2015outage, arezumand2017outage, varshney2017cognitive, varshney2018cognitive, al2017outage, erdogan2019error}. In \cite{arezumand2017outage}, the performance analysis was accomplished by developing analytical expressions of outage probability (OP) where RF-FSO links were subjected to Nakagami-$m$ fading and double generalized Gamma (DGG) turbulence, respectively. However, this model was extended in \cite{varshney2017cognitive,varshney2018cognitive} where the impact of antenna diversity in RF links was observed. In \cite{al2017outage}, the OP analysis was performed for amplify-and-forward (AF) relaying-based dual-hop network while multi-users and multi-destination were taken into consideration. The error probability (EP) expression was derived in \cite{erdogan2019error} where the influence of RF and FSO parameters on the system performance was inspected.

Security has become a widespread concern in the next-generation wireless communications (5G and beyond) due to its high risk of information leakage \cite{ibrahim2021enhancing,8500351}. In this domain, physical layer security (PLS) has been recognized as a complementary choice to the traditional cryptography strategy \cite{shakir2020physical, 9330523}. According to the comprehensive literature analysis, current PLS research has mainly limited on hybrid RF / FSO models \cite{shakir2020physical,ai2020secrecy,kafafy2020secure} and mixed RF-FSO networks \cite{sarker2020secrecy,lei2018secrecy,islam2020secrecy}. In \cite{shakir2020physical}, the PLS of a hybrid FSO / RF network was examined and it was found that secrecy performance is highly dependent on air turbulence, pointing error, and RF fading parameters. In \cite{ai2020secrecy}, authors obtained the secrecy diversity gain for RF backhaul system along with parallel FSO link whereas the impact of power allocation on secrecy performance was investigated in \cite{kafafy2020secure} for the parallel mmWave and FSO links. Including link blockage (LB) probability in the FSO link due to the sudden presence of moving objects (e.g. clouds, birds, insects, etc.), \cite{odeyemi2020secrecy} introduced the strategy of enhancing secrecy behavior for CR-based hybrid RF / FSO system.

\subsection{Motivation and Contributions}

Despite the fact that a CR-based hybrid RF / FSO system configuration is more robust and practical than a single RF or FSO link \cite{ai2020secrecy}, there are very few studies that have looked into the secrecy analysis of such systems and majority of the previously mentioned research focused on performance analysis of hybrid RF / FSO systems only. As a result, the investigation of PLS over the CR-based hybrid model is currently an open concept. To the best of authors' knowledge, there are no works that conduct secrecy analysis of a hybrid CR RF / FSO model considering double power constraints (i.e. the transmit power constraints at the $SU$ and maximum allowable interference power at the $PU$). Being motivated by this research gap, in this work, the security performance for an underlay CR based hybrid RF / FSO fading channel is demonstrated taking into account the single power constraint (i.e. maximum permissible interference at the $SU$) and the double power constrains in the presence of an eavesdropper that can intercept the secure communication via a RF link. The generalized $\alpha$-$\mu$ fading channel is considered for all the RF links since it can characterize small scale fading perfectly \cite{yacoub2007alpha}. The FSO link, on the other hand, is assumed to follow M\'alaga fading distribution, since it is also a generalized model and capable of accurately describing the pointing error and turbulence scenarios \cite{islam2021impact}. The contributions of this research are highlighted as follows:

\begin{enumerate}

\item Firstly, the expressions of cumulative density function (CDF) for selection combining (SC) based hybrid cognitive spectrum-sharing RF / FSO network assuming two different scenarios (e.g. single and double power constraints) are derived in closed-form. It is noteworthy that most of the works in the literature explored secrecy behaviour of the CR RF / FSO schemes considering maximum interference power constrain at $PU$ only, which is just a single part of the proposed model. In this perspective, considering both single and double power constraints in such type of hybrid model is certainly a novel and more practical approach. Moreover, this expressions of CDF also incorporates some other existing works \cite{odeyemi2019selection, shakir2019performance, shakir2020physical, ai2020secrecy, odeyemi2020secrecy} as special cases.

\item Secondly, the secrecy characteristics are derived for the proposed system model utilizing the underlay approach over the RF links and developing the analytical expressions for secrecy outage probability (SOP), probability of strictly positive secrecy capacity (SPSC), and effective secrecy throughout (EST) for the two scenarios. Since the considered $\alpha-\mu$ and M\'alaga models are generalized, and combination of the two in such type of hybrid model has been considered for the first time in the literature, according to the authors' knowledge, the derived expressions are also novel and generalized.

\item The derived expressions are further utilized to realize some numerical outcomes with some specific figures demonstrating impacts of each system parameters i.e. fading, primary-secondary interference, LB probability, pointing error, detection techniques, and various atmospheric turbulence conditions, etc. It is also notable that both detection techniques, i.e. heterodyne detection (HD) and intensity modulation/direct detection (IM/DD) techniques, have not been considered yet in such type of hybrid system model except in this work. Finally, the accuracy of the deduced analytical expressions are corroborated via Monte-Carlo (MC) simulations.

\end{enumerate}

\subsection{Organization}

The outline of this paper is arranged as follows: In section-\ref{system}, \ref{channel}, and \ref{cognitive}, the proposed framework along with the channel models is described. The novel analytical expressions of the performance metrics (i.e. SOP, probability of SPSC, and EST analysis for the two scenarios) are demonstrated in Section-\ref{metrics}. Section-\ref{results} incorporates some exclusive numerical results followed by insightful discussions. Finally, the concluding remarks of this work are presented in Section-\ref{conclusions}.

\section{System Model}
\label{system}

\begin{figure*}[!ht]
\vspace{0mm}
    \centerline{\includegraphics[width=0.9\textwidth,angle =0]{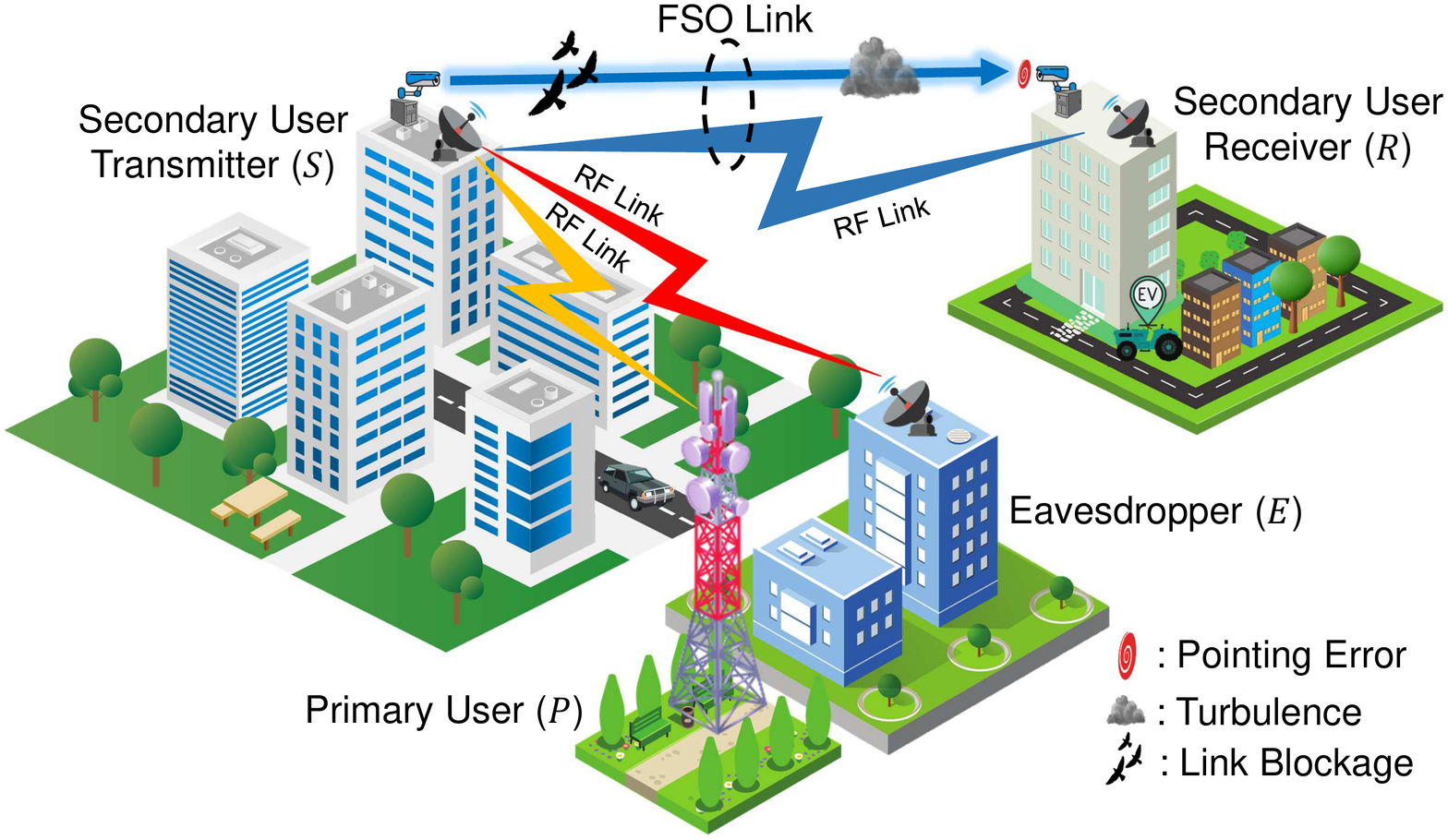}}
        \vspace{0mm}
 \caption{Proposed system model for hybrid RF / FSO based cognitive underlay network.}
    \label{f11}
\end{figure*}
A hybrid cognitive underlay network (CUN) is illustrated in Fig. \ref{f11}, which incorporates two parallel links: cognitive RF and FSO sub-systems, respectively. It is essential to mention that FSO allows the benefits of using a license-free spectrum with high data rates \cite{bag2018performance}. But in the FSO sub-system, the signal transmission can be substantially disrupted by atmospheric turbulence conditions and pointing errors. Hence, the RF sub-system is utilized to transmit the same information concurrently with the FSO sub-system thereby maintaining a continuous data transmission. To prevent any data loss, the receiver $R$ exploits selection combining diversity and always picks the best signal, i.e. the signal with highest (signal-to-noise ratio) SNR, among the two sub-links. Since the secondary user ($S$) utilizes the given licensed spectrum of the primary user ($P$), in the proposed underlay technique, the system requires that there must be no adverse interference at $P$.

\subsubsection{Cognitive RF Sub-link}

Based on the underlay approach, $S$ transmits its concealed information to $R$ under the following constraints:

\begin{itemize}

\item The peak tolerable interference power impinged by $S$ on $P$ cannot exceed a predefined value $P_{Q}$. In such a case, $S$ is not considered as a power limited terminal and it has the full freedom to utilize its power depending on a single power constraint, i.e. interference severity at $P$. This particular case is denoted as \textit{Scenario-I} in the rest of this work.

\item The CUN in \textit{Scenario-I} sometimes may not be practical, specifically when the channel coefficient of $S-P$ link rapidly fluctuates. This may lead to feedback burden since the instantaneous feedback gain is difficult to track. To mitigate these types of feedback burden and feedback errors, mean value power allocation strategy can be adopted \cite{afana2016quadrature}. Hence, besides only interference/single power constraint case, in double power constraint case, $S$ is assumed to be a power limited terminal and allowed to exploit maximum transmit power $P_{T}$. This particular case is denoted as \textit{Scenario-II} in the remaining manuscript.
    
\end{itemize}

\noindent
For \textit{Scenario-II}, utilizing mean value power allocation, the transmit power of $S$ is denoted as
\begin{align}
P_{t,II}= \min\biggl(\frac{P_{Q}}{|g_{p}|^2},P_{T} \biggl),
\end{align}
where $g_{p}$ denotes the channel gain of $S-P$ interference link. Therefore, the SNR at $R$ is expressed as
\begin{align}
\gamma_{r,II}=\min \biggl(\frac{\Psi_{Q}}{|g_{p}|^2},\Psi_{T} \biggl)|g_{r}|^2,
\end{align}
where $\Psi_{Q}=\frac{P_{Q}}{N_{r}}$, $\Psi_{T}=\frac{P_{T}}{N_{r}}$, $N_{r}$ represents the noise power imposed on $R$, and $g_{r}$ is the channel gain of $S-R$ RF sub-link. However, when $S$ is not a power-limited terminal (\textit{Scenario-I}), the transmit power at $S$ is represented as
\begin{align}
P_{t,I}=\frac{P_{Q}}{|g_{p}|^{2}}.
\end{align}
In this case, the SNR of $S-R$ RF sub-link is expressed as
\begin{align}
\gamma_{r,I}=\frac{\Psi_{Q}|g_{r}|^{2}}{|g_{p}|^{2}}.
\end{align}
\subsubsection{FSO Sub-link}

The FSO sub-system comprises of $S$ with a transmit aperture, FSO sub-link, and $R$ with a receive photo-detector. The transmitter dispatches information in optical form that is further converted into electrical one utilizing the photo-detector at $R$. Thereby, instantaneous SNR for $S-R$ FSO sub-link is written as
\begin{align}
\gamma_{o}=\frac{P_{f}}{N_{o}}\|g_{o}\|^{2},
\end{align}
where $g_{o}$ signifies the channel gain of $S-R$ FSO sub-link, $N_{o}$ denotes the optical noise power at $R$, and $P_{f}$ is the transmit optical power at $S$.

Finally, after selection combining at $R$, the received SNR is written as \cite[Eq.~(11)]{shakir2020physical}
\begin{align}
\label{eqn:ifsucdf12}
\gamma_{f,j}= \max(\gamma_{r,j},\gamma_{o}),
\end{align}
where $j \in \{I,II\}$.

\subsubsection{Eavesdropper Link}

In the proposed system, an eavesdropper ($E$) targets the RF sub-link to intercept the confidential data stream. The instantaneous SNR at $E$ is expressed as
\begin{align}
\gamma_{e,j}=\frac{P_{t,j}}{N_{e}}\|g_{e}\|^{2},
\end{align}
where $g_{e}$ denotes the channel gain of $S-E$ link and $N_{e}$ is the noise power at $E$.

\section{Channels Realization}
\label{channel}

In this section, the channel models of the $S-R$ (RF and FSO sub-links), $S-E$, and $S-P$ links are realized individually for the further utilization in the mathematical modeling of the proposed CUN system.

\subsection{PDF and CDF of SNRs for the RF Links}

The $S-R$ RF sub-link, $S-P$, and $S-E$ links are assumed to follow $\alpha-\mu$ distribution, where $\alpha$ denotes the non-linearity parameter of propagation environment ($\alpha>0$) and $\mu$ represents the number of multi-path clusters ($\mu>0$). Since $\alpha-\mu$ model exhibits generalized characteristics \cite{lei2017secrecyal}, several multi-path models \cite{4067122}, e.g. exponential, Rayleigh, Weibull, Gaussian, Gamma, and Nakagami-$m$, etc., can be obtained by setting different values of $\alpha$ and $\mu$. The PDF of $\gamma_{i}$, $i \in \{r, p, e\}$, where $r,p,$ and $e$ correspond to the $S-R$ (RF), $S-P$, and $S-E$ links, respectively, is given as \cite[Eq.~(2)]{kong2018secrecy}
\begin{align}
\label{eqn:rfallpdf}
f_{\gamma_{i}}(\gamma)=\frac{\alpha_{i}\delta_{i}^{\mu_{i}}}{2\Gamma(\mu_{i})}e^{-\delta_{i}\gamma^{\tilde{\alpha_{i}}}}\gamma^{\Theta_{i}}.
\end{align}
Here $\delta_{i}=\Phi_{i}^{-\tilde{\alpha_{i}}}$, $\Theta_{i}=\tilde{\alpha_{i}}\mu_{i}-1$, $\tilde{\alpha_{i}}=\frac{\alpha_{i}}{2}$, $\Phi_{i}$ is the average SNR of the RF links, and $\Gamma(\cdot)$ denotes the Gamma operator. The CDF of $\gamma_{i}$ is defined as
\begin{align}
\label{eqn:rfcdf}
F_{\gamma_{i}}(\gamma)=\int_{0}^{\gamma}f_{\gamma_{i}}(\gamma) d\gamma.
\end{align}
Plugging \eqref{eqn:rfallpdf} into \eqref{eqn:rfcdf}, and after performing integration upon utilizing \cite[Eq.~(3.381.8)]{GR:07:Book}, the CDF leads to
\begin{align}
\label{eqn:rfcdf3}
F_{\gamma_{i}}(\gamma)=\frac{\gamma(\mu_{i},\delta_{i}\gamma^{\tilde{\alpha_{i}}})}{\Gamma(\mu_{i})},
\end{align}
where $\gamma(,.,)$ denotes the lower incomplete Gamma function. Finally, with the help of \cite[Eq.~(8.352.6)]{GR:07:Book}, the CDF of $\gamma_{i}$ can be expressed in an alternative form as
\begin{align}\label{eqn:rfcdf4}
F_{\gamma_{i}}(\gamma)=1-e^{-\delta_{i}\gamma^{\tilde{\alpha_{i}}}}\sum_{m_{i}=0}^{\mu_{i}-1}\frac{(\delta_{i}\gamma^{\tilde{\alpha_{i}}})^{m_{i}}}{m_{i}!}.
\end{align}

\subsection{PDF and CDF of SNR for FSO SUB-Link}

The $S-R$ FSO sub-link is considered to follow M\'alaga ($\mathcal{M}$) turbulence. Since M\'alaga turbulence is a generalized FSO channel, it incorporates some other distributions such as Gamma-Gamma, lognormal, Rice-Nakagami, and Gamma as special cases \cite[Table~II]{islam2020secrecy}. Hence, the PDF of $\gamma_{o}$ is expressed as \cite[~Eq. (9)]{ansari2015performance}
\begin{align}
\label{eqn:rfcdf31}
f_{\gamma_{o}}(\gamma)=\frac{\epsilon^{2}\chi_{o}}{2^{s}\gamma}\sum_{m_{o}=0}^{\beta_{o}}\vartheta_{n}G_{1,3}^{3,0}\left[\varpi\biggl(\frac{\gamma}{\mu_{s}}\biggl)^\frac{1}{s}\biggl |
\begin{array}{c}
\epsilon ^2+1 \\
\epsilon ^2,\alpha_{o},m_{o} \\
\end{array}
\right],
\end{align}
where
\begin{align}
\nonumber
\chi_{o}&=\frac{2\alpha_{o}^{\frac{\alpha_{o}}{2}}}{g^{1+\frac{\alpha_{o}}{2}}\Gamma(\alpha_{o})}\biggl(\frac{g\beta_{o}}{g\beta_{o}+\Omega_{o}}\biggl)^{\beta_{o}+\frac{\alpha_{o}}{2}},
\\\nonumber
\varpi&=\frac{\epsilon^{2}\alpha_{o}\beta_{o}(g+\Omega_{o})}{(\epsilon^{2}+1)(g\beta_{o}+\Omega_{o})},
\\\nonumber
\vartheta_{n}&=\upsilon_{n}\biggl(\frac{\alpha_{o}\beta_{o}}{g\beta_{o}+\Omega_{o}}\biggl)^{-\frac{\alpha_{o}+m_{o}}{2}},
\\\nonumber
\upsilon_{n}&=\binom{\beta_{o}-1}{m_{o}-1}\frac{(g\beta_{o}+\Omega_{o})^{1-\frac{m_{o}}{2}}}{(m_{o}-1)!}\biggl(\frac{\Omega_{o}}{g}\biggl)^{m_{o}-1}\biggl(\frac{\alpha_{o}}{\beta_{o}}\biggl)^{\frac{m_{o}}{2}},
\end{align}
$\alpha_{o}$ and $\beta_{o}$ denote the fading parameters, $g=2b_{o}(1-\rho)$ represents the average power of scattering components received through off-axis eddies, $2b_{o}$ denotes the total average power of scattering components, $\epsilon$ ($0\leq\rho\leq1$) signifies the amount of scattering power coupled to the LOS component, the average power realized from coherent contributions is symbolized by $\Omega_{o}=\acute{\Omega_{o}}+2b_{o}\rho+2\sqrt{2b_{o}\rho\acute{\Omega_{o}}}\cos(\phi_{A}-\phi_{B})$, $\acute{\Omega_{o}}$ is termed as the average power of LOS components, $\phi_{A}$ and $\phi_{B}$ represent the deterministic phases of LOS, $\mu_{1}=\Phi_{o}$, $\mu_{2}=\frac{\alpha_{o}\epsilon^{2}(\epsilon^{2}+1)^{-2}(\epsilon^{2}+2)(g+\Omega_{o})\Phi_{o}}{(\alpha_{o}+1)[2g(g+2\Omega_{o})+\Omega_{o}^{2}(1+1/\beta_{o})]}$, $s$ denotes the detection technique (i.e. $s=1$ refers to HD technique and $s=2$ refers to IM/DD technique), $\Phi_{o}$ denotes the average SNR of FSO sub-link that is related to electrical SNR $\mu_{s}$, and G$[\cdot]$ represents the Meijer's G function \cite{Calculationbook02}. Plugging \eqref{eqn:rfcdf31} into \eqref{eqn:rfcdf}, the CDF of $\gamma_{o}$ is written as \cite[~Eq. 11]{ansari2015performance}
\begin{align}\label{eqn:cdffso}
F_{\gamma_{o}}(\gamma)=K\sum_{m_{o}=0}^{\beta_{o}}\varsigma_{n}G_{s+1,3s+1}^{3s,1}\left[\frac{V\gamma}{\mu_{s}}\biggl |
\begin{array}{c}
1,q_{1} \\
q_{2},0 \\
\end{array}
\right],
\end{align}
where
$K=\frac{\epsilon ^2\chi_{o}}{2^{s}(2\pi)^{s-1}}$, $\varsigma_{n}=\vartheta_{n}s^{\alpha_{o}+m_{o}-1}$, $V=\frac{\varpi^{s}}{s^{2s}}$,
$q_{1}=\{\frac{\epsilon^{2}+1}{s},\cdots,\frac{\epsilon^{2}+s}{s}\}$ including $s$ number of terms, and $q_{2}=\{\frac{\epsilon^{2}}{s},\cdots,\frac{\epsilon^{2}+s-1}{s},\frac{\alpha_{o}}{s},\cdots,\frac{\alpha_{o}+s-1}{s},\frac{m_{o}}{s},\cdots,\frac{m_{o}+s-1}{s}\}$ including $3s$ number of terms.

The FSO link can be temporarily blocked due to its extreme LOS requirements. Hence, including link blocking probability ($P_{o}$) in FSO and assuming SNR of the $S-R$ FSO sub-link as $\gamma_{o^{*}}$, the PDF of $\gamma_{o^{*}}$ is expressed finally as \cite[Eq.~(10)]{djordjevic2016outage}
\begin{align}
f_{\gamma_{o^{*}}}(\gamma)=P_{o}\delta(\gamma)+(1-P_{o})f_{\gamma_{o}}(\gamma),
\end{align}
where $\delta(\cdot)$ denotes a Dirac delta function \cite[Eq.~(14.03.02.0001.01)]{TheMathe2:online}. Now, the CDF of $\gamma_{o^{*}}$ is expressed as
\begin{align}\label{eqn:cdffsobk}
F_{\gamma_{o^{*}}}(\gamma)&=\int_{0}^{\gamma}f_{\gamma_{o^{*}}}(\gamma)d\gamma\cong P_{o}+(1-P_{o})F_{\gamma_{o}}(\gamma).
\end{align}
Substituting \eqref{eqn:cdffso} into \eqref{eqn:cdffsobk}, we obtain
\begin{align}\label{eqn:fso-bkf}
F_{\gamma_{o^{*}}}(\gamma)&=P_{o}+(1-P_{o})K\sum_{m_{o}=0}^{\beta_{o}}\varsigma_{n}\,G_{s+1,3s+1}^{3s,1}\left[\frac{V\gamma}{\mu_{s}}\biggl |
\begin{array}{c}
1,q_{1} \\
q_{2},0 \\
\end{array}
\right].
\end{align}

\section{Cognitive Underlay Network}
\label{cognitive}

This section demonstrates the CDFs of SNRs for the hybrid RF / FSO CUN considering both scenarios (Scenarios- I \& II).

\subsection{Scenario I}

Considering the interference of $S-P$ link and denoting SNR of the $S-R$ RF sub-link as $\gamma_{r,I}$, the CDF of $\gamma_{r,I}$ is written as \cite[Eq.~(4)]{erdogan2019error}
\begin{align}\label{eqn:ifsucdf}
F_{\gamma_{r,I}}(\gamma)=\int_{0}^{\infty}F_{\gamma_{r}}\biggl(\frac{\gamma}{\Psi_{Q}}x \biggl)f_{\gamma_{p}}(x)dx.
\end{align}
Now substituting \eqref{eqn:rfcdf4} and \eqref{eqn:rfallpdf} into \eqref{eqn:ifsucdf}, the CDF is expressed further as
\begin{align}
\nonumber
F_{\gamma_{r,I}}(\gamma)&=1-\sum_{m_{r}=0}^{\mu_{r}-1}\frac{\alpha_{p}\delta_{r}^{m_{r}}\delta_{p}^{\mu_{p}}}{2\Gamma(\mu_{p})m_{r}!}\biggl(\frac{\gamma}{\Psi_{Q}}\biggl)^{\tilde{\alpha_{r}}m_{r}}
\\
&\times\int_{0}^{\infty} e^{-(\delta_{p}x^{\tilde{\alpha_{p}}}+\Xi_{1}^{I}\gamma_{r}^{\tilde{\alpha_{r}}}x^{\tilde{\alpha_{r}}})}x^{\Theta_{p}+\tilde{\alpha_{r}}m_{r}} dx,
\end{align}
where $\Xi_{1}^{I}=\delta_{r}\Psi_{Q}^{-\tilde{\alpha_{r}}}$. Now, following some algebraic manipulations, assuming $\tilde{\alpha_{p}}=\tilde{\alpha_{r}}$, and performing integration utilizing \cite[~Eq. (3.326.2)]{GR:07:Book}, the CDF of $S-R$ RF sub-link is finally obtained as
\begin{align}\label{eqn:rfunderlay}
\nonumber
F_{\gamma_{r,I}}(\gamma)&=1-\sum_{m_{r}=0}^{\mu_{r}-1}\frac{\Gamma(\Xi_{2}^{I})\alpha_{p}\delta_{r            }^{m_{r}}\delta_{p}^{\mu_{p}}}{\tilde{\alpha_{p}}2\Gamma(\mu_{p})m_{r}!}\biggl(\frac{\gamma}{\Psi_{Q}}\biggl)^{\tilde{\alpha_{r}}m_{r}}
\\
&\times(\Xi_{1}^{I}\gamma^{\tilde{\alpha_{r}}}+\Phi_{p}^{-\tilde{\alpha_{p}}})^{-\Xi_{2}^{I}},
\end{align}
where $\Xi_{2}^{I}=\frac{\tilde{\alpha_{r}}m_{r}+\mu_{p}\tilde{\alpha_{p}}}{\tilde{\alpha_{p}}}$. According to \eqref{eqn:ifsucdf12}, the CDF of end-to-end SNR at $R$ is written as
\begin{align}
\label{eqn:rq1}
F_{\gamma_{f,I}}(\gamma)=F_{\gamma_{r,I}}(\gamma)\times F_{\gamma_{o^{*}}}(\gamma).
\end{align}
By replacing \eqref{eqn:rfunderlay} and \eqref{eqn:fso-bkf} into \eqref{eqn:rq1}, the CDF of $\gamma_{f,I}$ is derived in \eqref{eqn:EQv}.
\begin{figure*}
\begin{align}
\label{eqn:EQv}
\nonumber
F_{\gamma_{f,I}}(\gamma)&=P_{o}+(1-P_{o})\sum_{m_{o}=0}^{\beta_{o}}K\varsigma_{n}
G_{s+1,3s+1}^{3s,1}\left[\frac{V\gamma}{\mu_{s}}\biggl |
\begin{array}{c}
1,q_{1} \\
q_{2},0 \\
\end{array}
\right]
-P_{o}\sum_{m_{r}=0}^{\mu_{r}-1}\frac{\delta_{r}^{m_{r}}\alpha_{p}\delta_{p}^{\mu_{p}}\Gamma(\Xi_{2}^{I})}{2\Gamma(\mu_{p})\tilde{\alpha_{p}}m_{r}!}
\biggl(\frac{\gamma}{\Psi_{Q}}\biggl)^{\tilde{\alpha_{r}}m_{r}}(\Xi_{1}^{I}\gamma^{\tilde{\alpha_{p}}}+\Phi_{p}^{-\tilde{\alpha_{p}}})^{-\Xi_{2}^{I}}
\\
&-(1-P_{o}) \sum_{m_{r}=0}^{\mu_{r}-1}\sum_{m_{o}=0}^{\beta_{o}}K\varsigma_{n}\frac{\Gamma(\Xi_{2}^{I})\alpha_{p}\delta_{p}^{\mu_{p}}\delta_{r}^{m_{r}}}{\tilde{\alpha_{p}}2\Gamma(\mu_{p})m_{r}!}\biggl(\frac{\gamma}{\Psi_{Q}}\biggl)^{\tilde{\alpha_{r}}m_{r}}
(\Xi_{1}^{I}\gamma^{\tilde{\alpha_{p}}}+\Phi_{p}^{-\tilde{\alpha_{p}}})^{-\Xi_{2}^{I}}
G_{s+1,3s+1}^{3s,1}\left[\frac{V\gamma}{\mu_{s}}\biggl |
\begin{array}{c}
1,q_{1} \\
q_{2},0 \\
\end{array}
\right].
\end{align}
\hrulefill
\end{figure*}

\subsection{Scenario II}
Assuming double power constraints in the RF sub-link, the CDF of $ F_{\gamma_{r,II}}$ is written as
\begin{align}
\label{eqn:frf}
\nonumber
F_{\gamma_{r,II}}(\gamma_{r})&=\Pr\biggl \{\min \biggl(\frac{\Psi_{Q}}{|g_{p}|^2},\Psi_{T} \biggl)|g_{r}|^2 \leq \gamma_{r}\biggl \}
\\
\nonumber
&= \underbrace{\Pr\biggl \{|g_{r}|^2 \leq \frac{\gamma_{r}}{\Psi_{T}},\frac{\Psi_{Q}}{|g_{p}|^2}  \geq \Psi_{T} \biggl \}}_{\Lambda_{1}}
\\
& + \underbrace{\Pr\biggl \{\frac{|g_{r}|^2}{|g_{p}|^2} \leq \frac{\gamma_{r}}{\Psi_{Q}},\frac{\Psi_{Q}}{|g_{p}|^2} \leq \Psi_{T}\biggl \}}_{\Lambda_{2}}.
\end{align}
Since $g_{r}$ and $g_{p}$ are completely independent to one other, hence, $\Lambda_{1}$ is expressed in an alternative form as
\begin{align} \nonumber \label{eqn:a1}
\Lambda_{1}&= \Pr \biggl\{|g_{p}|^2 \leq \frac{\Psi_{Q}}{\Psi_{T}}\biggl\} \Pr \biggl\{|g_{r}|^2 \leq \frac{\gamma_{r}}{\Psi_{T}} \biggl\}
\\
&= F_{\gamma_{p}}\biggl(\frac{\Psi_{Q}}{\Psi_{T}} \biggl) F_{\gamma_{r}}\biggl(\frac{\gamma}{\Psi_{T}} \biggl).
\end{align}
Substituting \eqref{eqn:rfcdf4} into \eqref{eqn:a1}, $\Lambda_{1}$ can be finally written as
\begin{align} \label{eqn:a11}
\nonumber
\Lambda_{1}&=1-\sum_{m_{p}=0}^{\mu_{p}-1}\Xi_{1}^{II}-\sum_{m_{r}=0}^{\mu_{r}-1}\Xi_{2}^{II}\gamma^{\tilde{\alpha_{r}}m_{r}}e^{-\delta_{r}\Psi_{T}^{-\tilde{\alpha_{r}}}\gamma^{\tilde{\alpha_{r}}}}
\\
&+\sum_{m_{p}=0}^{\mu_{p}-1}\sum_{m_{r}=0}^{\mu_{r}-1}\Xi_{3}^{II}\gamma^{\tilde{\alpha_{r}}m_{r}}e^{-(\delta_{p}\Psi_{Q}^{\tilde{\alpha_{p}}}\Psi_{T}^{-\tilde{\alpha_{p}}}+\delta_{r}\Psi_{T}^{-\tilde{\alpha_{r}}}\gamma^{\tilde{\alpha_{r}}})},
\end{align}
where $\Xi_{1}^{II}=\frac{\delta_{p}^{m_{p}}}{m_{p}!}\biggl(\frac{\Psi_{Q}}{\Psi_{T}}\biggl)^{\tilde{\alpha_{p}}m_{p}}e^{-\delta_{p}\Psi_{Q}^{\tilde{\alpha_{p}}}\Psi_{T}^{-\tilde{\alpha_{p}}}}$, $\Xi_{2}^{II}=\frac{\delta_{r}^{m_{r}}}{m_{r}!}\Psi_{T}^{-\tilde{\alpha_{r}}m_{r}}$, and $\Xi_{3}^{II}=\frac{\delta_{p}^{m_{p}}\delta_{r}^{m_{r}}}{m_{r}!m_{p}!}\left(\frac{\Psi_{Q}}{\Psi_{T}}\right)^{\tilde{\alpha_{p}}m_{p}}\Psi_{T}^{-\tilde{\alpha_{r}}m_{r}}$. Now, according to the concept of probability theory \cite{papoulis1989probability}, $\Lambda_{2}$ is written as
\begin{align} \nonumber \label{eqn:ai}
\Lambda_{2}&= \int_{\frac{\Psi_{Q}}{\Psi_{T}}}^{\infty}f_{\gamma_{p}}(y)\int_{0}^{\frac{\gamma_{r}}{\Psi_{Q}}y}f_{\gamma_{r}}(x) dx dy
\\
&= \int_{\frac{\Psi_{Q}}{\Psi_{T}}}^{\infty}f_{\gamma_{p}}(y) F_{\gamma_{r}} \biggl(\frac{\gamma y}{\Psi_{Q}} \biggl) dy.
\end{align}
Replacing \eqref{eqn:rfallpdf} and \eqref{eqn:rfcdf4} into \eqref{eqn:ai}, $\Lambda_{2}$ is obtained as
\begin{align} \nonumber
\label{eqn:a22}
\Lambda_{2}&=\Xi_{5}^{II}-\sum_{m_{r}=0}^{\mu_{r}-1}\sum_{m_{3}=0}^{\Omega-1}\sum_{m_{4}=0}^{m_{3}}\sum_{m_{5}=0}^{\infty}\binom{m_{3}}{m_{4}}\binom{\Omega+m_{5}-1}{m_{5}}
\\
&\times \Xi_{6}^{II} \gamma^{\tilde{\alpha_{r}}(m_{r}+m_{4}+m_{5})}e^{-(\delta_{p}\Psi_{Q}^{\tilde{\alpha_{r}}}\Psi_{T}^{-\tilde{\alpha_{r}}}+\delta_{r}\Psi_{T}^{-2\tilde{\alpha_{r}}}\Psi_{Q}^{-\tilde{\alpha_{r}}}\gamma^{\tilde{\alpha_{r}}})},
\end{align}
where $\Xi_{5}^{II}=\frac{\alpha_{p}\delta_{p}^{\mu_{p}}}{2 \Gamma(\mu_{p})\tilde{\alpha_{p}}\delta_{p}^{\Xi_{4}^{II}}} \Gamma \left[\Xi_{4}^{II},\delta_{p}\left(\frac{\Psi_{Q}}{\Psi_{T}}\right)^{\tilde{\alpha_{p}}}\right]$, $\Xi_{4}^{II}=\frac{\Theta_{p}+1}{\tilde{\alpha_{p}}}$, $\Omega=\frac{\Theta_{p}+\tilde{\alpha_{r}}m_{r}+1}{\tilde{\alpha_{r}}}$, and $\Xi_{6}^{II}= \frac{\alpha_{p}\delta_{r}^{m_{r}+m_{4}+m_{5}} (\Omega-1)!(-1)^{m_{5}}}{2\Gamma(\mu_{p})m_{r}!\tilde{\alpha_{r}}m_{3}!} \delta_{p}^{-(\Omega+m_{5}-m_{3}+m_{4}-\mu_{p})}
\\
\Psi_{Q}^{-\tilde{\alpha_{r}}(m_{r}-m_{3}+m_{5})}
(\Psi_{T}^{-\tilde{\alpha_{r}}})^{m_{3}+m_{4}}$. For proof, please see the Appendix. Finally, utilizing \eqref{eqn:a11} and \eqref{eqn:a22} into \eqref{eqn:frf}, and performing some simple mathematical operations, $F_{\gamma_{r,II}}(\gamma)$ is written as shown in \eqref{eqn:frf12}.
\begin{figure*}[!t]
\begin{align}
\label{eqn:frf12}
\nonumber
F_{\gamma_{r,II}}(\gamma)&=1+\Xi_{5}^{II}-\sum_{m_{p}=0}^{\mu_{p}-1}\Xi_{1}^{II}-\sum_{m_{r}=0}^{\mu_{r}-1}\Xi_{2}^{II}\gamma^{\tilde{\alpha_{r}}m_{r}}e^{-\delta_{r}\Psi_{T}^{-\tilde{\alpha_{r}}}\gamma^{\tilde{\alpha_{r}}}}+\sum_{m_{p}=0}^{\mu_{p}-1}\sum_{m_{r}=0}^{\mu_{r}-1}\Xi_{3}^{II}\gamma^{\tilde{\alpha_{r}}m_{r}}e^{-(\delta_{p}\Psi_{Q}^{\tilde{\alpha_{p}}}\Psi_{T}^{-\tilde{\alpha_{p}}}+\delta_{r}\Psi_{T}^{-\tilde{\alpha_{r}}}\gamma^{\tilde{\alpha_{r}}})}
\\
&-\sum_{m_{r}=0}^{\mu_{r}-1}\sum_{m_{3}=0}^{\Omega-1}\sum_{m_{4}=0}^{m_{3}}\sum_{m_{5}=0}^{\infty}\binom{m_{3}}{m_{4}}\binom{\Omega+m_{5}-1}{m_{5}}\Xi_{6}^{II} \gamma^{\tilde{\alpha_{r}}(m_{r}+m_{4}+m_{5})}e^{-(\delta_{p}\Psi_{Q}^{\tilde{\alpha_{r}}}\Psi_{T}^{-\tilde{\alpha_{r}}}+\delta_{r}\Psi_{T}^{-2\tilde{\alpha_{r}}}\Psi_{Q}^{\tilde{\alpha_{r}}}\gamma^{\tilde{\alpha_{r}}})}.
\end{align}
\hrulefill
\end{figure*}
Now, similar to \eqref{eqn:rq1}, the CDF of $\gamma_{f,II}$ is expressed as
\begin{align} \label{eqn:ceq}
F_{\gamma_{f,II}}(\gamma)=F_{\gamma_{r,II}}(\gamma)\times F_{\gamma_{o^{*}}}(\gamma).
\end{align}
Now, substituting \eqref{eqn:frf12} and \eqref{eqn:fso-bkf} into \eqref{eqn:ceq}, the CDF of $\gamma_{f,II}$ is obtained as shown in \eqref{eqn:dcdf}, where $\mathcal{X}=1+\Xi_{5}^{II}-\sum_{m_{p}=0}^{\mu_{p}-1}\Xi_{1}^{II}$.

\begin{figure*}[!t]
\begin{align}
\label{eqn:dcdf}
\nonumber
F_{\gamma_{f,II}}(\gamma)&=P_{o}\Biggl\{\mathcal{X}-\sum_{m_{r}=0}^{\mu_{r}-1}\Xi_{2}^{II}\gamma^{\tilde{\alpha_{r}}m_{r}}e^{-\delta_{r}\Psi_{T}^{-\tilde{\alpha_{r}}}\gamma^{\tilde{\alpha_{r}}}}+\sum_{m_{p}=0}^{\mu_{p}-1}\sum_{m_{r}=0}^{\mu_{r}-1}\Xi_{3}^{II}\gamma^{\tilde{\alpha_{r}}m_{r}}e^{-(\delta_{p}\Psi_{Q}^{\tilde{\alpha_{p}}}\Psi_{T}^{-\tilde{\alpha_{p}}}+\delta_{r}\Psi_{T}^{-\tilde{\alpha_{r}}}\gamma^{\tilde{\alpha_{r}}})}-\sum_{m_{r}=0}^{\mu_{r}-1}\sum_{m_{3}=0}^{\Omega-1}
\\
\nonumber
&\times\sum_{m_{4}=0}^{m_{3}}\sum_{m_{5}=0}^{\infty}\binom{m_{3}}{m_{4}}\binom{\Omega+m_{5}-1}{m_{5}}\Xi_{6}^{II} \gamma^{\tilde{\alpha_{r}}(m_{r}+m_{4}+m_{5})}e^{-(\delta_{p}\Psi_{Q}^{\tilde{\alpha_{r}}}\Psi_{T}^{-\tilde{\alpha_{r}}}+\delta_{r}\Psi_{T}^{-2\tilde{\alpha_{r}}}\Psi_{Q}^{\tilde{\alpha_{r}}}\gamma^{\tilde{\alpha_{r}}})}\Biggl\}+(1-P_{o})K\sum_{m_{o}=0}^{\beta_{o}}\varsigma_{n}
\\
\nonumber
&\times G_{s+1,3s+1}^{3s,1}\left[\frac{V\gamma}{\mu_{s}}\biggl |
\begin{array}{c}
1,q_{1} \\
q_{2},0 \\
\end{array}
\right]\Biggl\{\mathcal{X}-\sum_{m_{r}=0}^{\mu_{r}-1}\Xi_{2}^{II}\gamma^{\tilde{\alpha_{r}}m_{r}}e^{-\delta_{r}\Psi_{T}^{-\tilde{\alpha_{r}}}\gamma^{\tilde{\alpha_{r}}}}+\sum_{m_{p}=0}^{\mu_{p}-1}\sum_{m_{r}=0}^{\mu_{r}-1} \Xi_{3}^{II}\gamma^{\tilde{\alpha_{r}}m_{r}}e^{-(\delta_{p}\Psi_{Q}^{\tilde{\alpha_{p}}}\Psi_{T}^{-\tilde{\alpha_{p}}}+\delta_{r}\Psi_{T}^{-\tilde{\alpha_{r}}}\gamma^{\tilde{\alpha_{r}}})}
\\
&-\sum_{m_{r}=0}^{\mu_{r}-1}\sum_{m_{3}=0}^{\Omega-1}\sum_{m_{4}=0}^{m_{3}}\sum_{m_{5}=0}^{\infty}\binom{m_{3}}{m_{4}}\binom{\Omega+m_{5}-1}{m_{5}}\Xi_{6}^{II}\gamma^{\tilde{\alpha_{r}}(m_{r}+m_{4}+m_{5})}\,e^{-(\delta_{p}\Psi_{Q}^{\tilde{\alpha_{r}}}\Psi_{T}^{-\tilde{\alpha_{r}}}+\delta_{r}\Psi_{T}^{-2\tilde{\alpha_{r}}}\Psi_{Q}^{\tilde{\alpha_{r}}}\gamma^{\tilde{\alpha_{r}}})}\Biggl\}.
\end{align}
\hrulefill
\end{figure*}

\section{Performance Analysis}
\label{metrics}

In this section, we demonstrate the novel analytical expressions for SOP, probability of SPSC, and EST utilizing \eqref{eqn:rfallpdf}, \eqref{eqn:EQv}, and \eqref{eqn:dcdf}.

\subsection{Secure Outage Probability Analysis}

In order to investigate the secrecy behaviour of the proposed hybrid system, SOP is one of the fundamental performance metrics. SOP particularly depends on whether the target secrecy rate ($\Upsilon_{e}$) is greater than secrecy capacity ($C_{s}$), where $C_{s}=[\log_{2}(1+\gamma_{f,j})-\log_{2}(1+\gamma_{e,j})]^{+}$ and $[z]^{+}=$ max$\{z,0\}$. So, mathematically, SOP can be described as \cite[Eq.~(14)]{[49]lei2016}
\begin{align}
\nonumber
SOP&= \Pr\left\{C_{s}(\gamma_{f,j},\gamma_{e,j})\leq \Upsilon_{e}\right\}
\\
\label{eqn:sopup}
&=\int_{0}^{\infty}F_{\gamma_{f,j}}(\sigma\gamma+\sigma-1)f_{\gamma_{e,j}}(\gamma)d\gamma,
\end{align}
where $\sigma=2^{\Upsilon_{e}}$ and $\Upsilon_{e}>0$. Since Meijer's $G$ function exists in  \eqref{eqn:EQv} and \eqref{eqn:dcdf}, derivation of SOP utilizing \eqref{eqn:sopup} is almost impossible. Hence, the lower bound of SOP can be derived as
\begin{align}\label{eqn:sop1}
SOP\geq SOP_{L}&=\int_{0}^{\infty}F_{\gamma_{f,j}}(\sigma\gamma)f_{\gamma_{e,j}}(\gamma)d\gamma.
\end{align}

\subsubsection{Scenario I}
Substituting \eqref{eqn:EQv} and \eqref{eqn:rfallpdf} into \eqref{eqn:sop1}, SOP is written as
\begin{align}
\label{eqn:sopf1}
\nonumber
SOP_{L}^{I}&=\frac{\alpha_{e}\delta_{e}^{\mu_{e}}P_{o}}{2\Gamma(\mu_{e})}\Im_{1}+(1-P_{o})K\varsigma_{n}\\
&\times\sum_{m_{o}=0}^{\beta_{o}}\biggl \{\frac{\alpha_{e}\delta_{e}^{\mu_{e}}}{2\Gamma(\mu_{e})} \Im_{2}-\Xi_{5}^{I} \Im_{4}\biggl \}-P_{o}\sum_{m_{r}=0}^{\mu_{r}-1}\Xi_{5}^{I}\Im_{3},
\end{align}
where $\Xi_{5}^{I}=\binom{\Xi_{2}^{I}+m_{2}-1}{m_{2}}\frac{\alpha_{p}\delta_{p}^{\mu_{p}}}{2\Gamma(\mu_{p})}\frac{\alpha_{e}\delta_{e}^{\mu_{e}}}{2\Gamma(\mu_{e})}\frac{\delta_{r}^{m_{r}}}{m_{r}!}\frac{\Gamma(\Xi_{2}^{I})}{\tilde{\alpha_{p}}}\left(\frac{\sigma}{\Psi_{Q}}\right)^{\tilde{\alpha_{r}}m_{r}}
\\
(-1)^{m_{2}}\left(\Xi_{1}^{I}\sigma^{\tilde{\alpha_{r}}}\right)^{m_{2}}\left(\Phi_{p}^{-\tilde{\alpha_{p}}}\right)^{-\Xi_{2}^{I}-m_{2}}$ and the four integral terms, $\Im_{1}$, $\Im_{2}$, $\Im_{3}$, and $\Im_{4}$ are derived as follows.

\subsubsection*{Derivation of $\Im_{1}$}

Utilizing the formula of  \cite[~Eq. (3.326.2)]{GR:07:Book}, $\Im_{1}$ is expressed as
\begin{align}\label{eqn:s1}
\Im_{1}=\int_{0}^{\infty}e^{-\delta_{e}\gamma^{\tilde{\alpha_{e}}}}\gamma^{\Theta_{e}}d\gamma=\frac{\Gamma\left(\frac{\Theta_{e}+1}{\tilde{\alpha_{e}}}\right)}{\tilde{\alpha_{e}}}\delta_{e}^{-\frac{\Theta_{e}+1}{\tilde{\alpha_{e}}}}.
\end{align}

\subsubsection*{Derivation of $\Im_{2}$}

$\Im_{2}$ is written as
\begin{align}
\Im_{2}=&\int_{0}^{\infty}e^{-\delta_{e}\gamma^{\tilde{\alpha_{e}}}}\gamma^{\Theta_{e}}G_{s+1,3s+1}^{3s,1}\left[\frac{V\sigma\gamma}{\mu_{s}}\biggl |
\begin{array}{c}
1,q_{1} \\
q_{2},0 \\
\end{array}
\right]d\gamma.
\end{align}
Now, with the help of \cite[~Eqs. (2.24.1.1) and (8.4.3.1)]{Calculationbook02}, $\Im_{2}$ is converted as
\begin{align}
\nonumber
\Im_{2}&=\int_{0}^{\infty}\gamma^{\Theta_{e}}
G_{0,1}^{1,0}\left[\delta_{e}\gamma^{\tilde{\alpha_{e}}}\biggl |
\begin{array}{c}
- \\
0 \\
\end{array}
\right]G_{s+1,3s+1}^{3s,1}\left[\frac{V\sigma\gamma}{\mu_{s}}\biggl |
\begin{array}{c}
1,q_{1} \\
q_{2},0 \\
\end{array}
\right] d\gamma
\\
\nonumber
&=\frac{\tilde{\alpha_{e}}^{s(2\Theta_{e}+1)}\left(\frac{V\sigma}{\mu_{s}}\right)^{-(\Theta_{e}+1)}}{(2\pi)^{s(\tilde{\alpha_{e}}-1)}}
\\
&\times G_{\tilde{\alpha_{e}}(3s+1),1+\tilde{\alpha_{e}}(s+1)}^{1+\tilde{\alpha_{e}}, \tilde{3s\alpha_{e}}}\left[\left.\frac{\delta_{e}\tilde{\alpha_{e}}^{2s\tilde{\alpha_{e}}}}{\left(\frac{V\sigma}{\mu_{s}}\right)^{\tilde{\alpha_{e}}}}\right|
\begin{array}{c}
\zeta_{1}, \Delta(\tilde{\alpha_{e}},-\Theta_{e}) \\
0, \zeta_{2}, \zeta_{3} \\
\end{array}
\right],
\end{align}
where $\zeta_{1}= \Delta(\tilde{\alpha_{e}},-\Theta_{e}-q_{2})$, $\zeta_{2}= \Delta(\tilde{\alpha_{e}},-\Theta_{e}-1)$, $\zeta_{3}= \Delta(\tilde{\alpha_{e}},-\Theta_{e}-q_{1})$, and $\Delta(c,r)=\frac{r}{c},\frac{r+1}{c}, ...  ,\frac{r+c-1}{c}$ as defined in \cite[Eq.~(22)]{adamchik1990algorithm}.

\subsubsection*{Derivation of $\Im_{3}$}

$\Im_{3}$ is expressed as
\begin{align}
\Im_{3}=&\int_{0}^{\infty}(\Xi_{1}^{I}\sigma^{\tilde{\alpha_{r}}}\gamma^{\tilde{\alpha_{p}}}+\Phi_{p}^{-\tilde{\alpha_{p}}})^{-\Xi_{2}^{I}}\,\gamma^{\Theta_{e}+\tilde{\alpha_{r}}m_{r}} e^{-\delta_{e}\gamma^{\tilde{\alpha_{e}}}}d\gamma.
\end{align}
Utilizing binomial expansion and \cite[~Eq. (3.326.2)]{GR:07:Book}, $\Im_{3}$ is expressed in alternative form as
\begin{align}
\Im_{3}=\sum_{m_{2}=0}^{\infty}\int_{0}^{\infty}\gamma^{\Xi_{3}^{I}}e^{-\delta_{e}\gamma^{\tilde{\alpha_{e}}}}d\gamma=\sum_{m_{2}=0}^{\infty}\frac{\Gamma(\Xi_{4}^{I})}{\tilde{\alpha_{e}}}\delta_{e}^{-\Xi_{4}^{I}},
\end{align}
where $\Xi_{3}^{I}=\Theta_{e}+\tilde{\alpha_{r}}m_{r}+\tilde{\alpha_{p}}m_{2}$ and $\Xi_{4}^{I}=\frac{\Xi_{3}^{I}+1}{\tilde{\alpha_{e}}}$.

\subsubsection*{Derivation of $\Im_{4}$}

Following the similar approach of $\Im_{2}$ and $\Im_{3}$, $\Im_{4}$ is illustrated as
\begin{align}
\nonumber
\Im_{4}&=\sum_{m_{2}=0}^{\infty} \int_{0}^{\infty}e^{-\delta_{e}\gamma^{\tilde{\alpha_{e}}}}\gamma^{\Xi_{3}^{I}} G_{s+1,3s+1}^{3s,1}\left[\frac{V\sigma\gamma}{\mu_{s}}\biggl |
\begin{array}{c}
1,q_{1} \\
q_{2},0 \\
\end{array}
\right] d\gamma
\\
\nonumber
&=\sum_{m_{2}=0}^{\infty}\frac{\tilde{\alpha_{e}}^{s(2\,\Xi_{3}^{I}+1)}\left(\frac{V\sigma}{\mu_{s}}\right)^{-(\Xi_{3}^{I}+1)}}{(2\pi)^{s(\tilde{\alpha_{e}}-1)}}
\\
&\times G_{\tilde{\alpha_{e}}(3s+1),1+\tilde{\alpha_{e}}(s+1)}^{1+\tilde{\alpha_{e}}, \tilde{3s\alpha_{e}}}\left[\frac{\delta_{e}\tilde{\alpha_{e}}^{2s\tilde{\alpha_{e}}}}{\left(\frac{V\sigma}{\mu_{s}}\right)^{\tilde{\alpha_{e}}}} \Biggl|
\begin{array}{c}
\zeta_{4}, \Delta(\tilde{\alpha_{e}},-\Xi_{3}^{I}) \\
0, \zeta_{5}, \zeta_{6} \\
\end{array}
\right].
\end{align}
where $\zeta_{4}= \Delta(\tilde{\alpha_{e}},-\Xi_{3}^{I}-q_{2})$, $\zeta_{5}= \Delta(\tilde{\alpha_{e}},-\Xi_{3}^{I}-1)$, and $\zeta_{6}= \Delta(\tilde{\alpha_{e}},-\Xi_{3}^{I}-q_{1})$.

\subsubsection{Scenario II}

Substituting \eqref{eqn:rfallpdf} and \eqref{eqn:dcdf} into \eqref{eqn:sop1}, SOP for double power constraints scenario is derived as \eqref{eqn:sopf}, where the eight integral terms, $\mathcal{R}_{1}$, $\mathcal{R}_{2}$, $\mathcal{R}_{3}$, $\mathcal{R}_{4}$, $\mathcal{R}_{5}$, $\mathcal{R}_{6}$, $\mathcal{R}_{7}$, and $\mathcal{R}_{8}$ are derived as follows.

\begin{figure*}[!t]
\begin{align}
\label{eqn:sopf}
\nonumber
&SOP_{L}^{II}=\frac{\alpha_{e}\delta_{e}^{\mu_{e}}}{2\Gamma(\mu_{e})}\Biggl[P_{o}\Biggl\{\mathcal{X}\mathcal{R}_{1}-\sum_{m_{r}=0}^{\mu_{r}-1}\Xi_{2}^{II}\sigma^{\tilde{\alpha_{r}}m_{r}}\mathcal{R}_{2}+\sum_{m_{p}=0}^{\mu_{p}-1}\sum_{m_{r}=0}^{\mu_{r}-1}\Xi_{3}^{II}\sigma^{\tilde{\alpha_{r}}m_{r}}e^{-\delta_{p}\Psi_{Q}^{\tilde{\alpha_{p}}}\Psi_{T}^{-\tilde{\alpha_{p}}}}\mathcal{R}_{3}-\sum_{m_{r}=0}^{\mu_{r}-1}\sum_{m_{3}=0}^{\Omega-1}\sum_{m_{4}=0}^{m_{3}}\sum_{m_{5}=0}^{\infty}
\\
\nonumber
&\times \binom{\Omega+m_{5}-1}{m_{5}}\binom{m_{3}}{m_{4}} \Xi_{6}^{II} \sigma^{\tilde{\alpha_{r}}(m_{r}+m_{4}+m_{5})}e^{-\delta_{p}\Psi_{Q}^{\tilde{\alpha_{r}}}\Psi_{T}^{-\tilde{\alpha_{r}}}}\mathcal{R}_{4}\Biggl\}+(1-P_{o})K\sum_{m_{o}=0}^{\beta_{o}}\varsigma_{n}\Biggl\{\mathcal{X}\mathcal{R}_{5}-\sum_{m_{r}=0}^{\mu_{r}-1}\Xi_{2}^{II}\mathcal{R}_{6}\,\sigma^{\tilde{\alpha_{r}}m_{r}}+\sum_{m_{p}=0}^{\mu_{p}-1}
\\
&\times \sum_{m_{r}=0}^{\mu_{r}-1} \Xi_{3}^{II}\mathcal{R}_{7}\sigma^{\tilde{\alpha_{r}}m_{r}}e^{-\delta_{p}\Psi_{Q}^{\tilde{\alpha_{p}}}\Psi_{T}^{-\tilde{\alpha_{p}}}}-\sum_{m_{r}=0}^{\mu_{r}-1}\sum_{m_{3}=0}^{\Omega-1}\sum_{m_{4}=0}^{m_{3}}\sum_{m_{5}=0}^{\infty}\binom{m_{3}}{m_{4}}\binom{\Omega+m_{5}-1}{m_{5}}\Xi_{6}^{II} \mathcal{R}_{8}\,\sigma^{\tilde{\alpha_{r}}(m_{r}+m_{4}+m_{5})}e^{-\delta_{p}\Psi_{Q}^{\tilde{\alpha_{r}}}\Psi_{T}^{-\tilde{\alpha_{r}}}}\Biggl\}\Biggl].
\end{align}
\hrulefill
\end{figure*}

\subsubsection*{Derivation of $\mathcal{R}_{1}$}
Utilizing \cite[~Eq. (3.326.2)]{GR:07:Book}, $\mathcal{R}_{1}$ is expressed as
\begin{align}
\mathcal{R}_{1}&=\int_{0}^{\infty}e^{-\delta_{e}\gamma^{\tilde{\alpha_{e}}}}\gamma^{\Theta_{e}}d\gamma=\frac{\Gamma\left(\frac{\Theta_{e}+1}{\tilde{\alpha_{e}}}\right)}{\tilde{\alpha_{e}}}\delta_{e}^{-\frac{\Theta_{e}+1}{\tilde{\alpha_{e}}}}.
\end{align}

\subsubsection*{Derivation of $\mathcal{R}_{2}$ and $\mathcal{R}_{3}$}
$\mathcal{R}_{2}$ and $\mathcal{R}_{3}$ are expressed as
\begin{align}
\mathcal{R}_{2}=\mathcal{R}_{3}=\int_{0}^{\infty}\gamma^{\tilde{\alpha_{r}}m_{r}+\Theta_{e}}e^{-\delta_{r}\sigma^{\tilde{\alpha_{r}}}\Psi_{T}^{-\tilde{\alpha_{r}}}\gamma^{\tilde{\alpha_{r}}}}e^{-\delta_{e}\gamma^{\tilde{\alpha_{e}}}}d\gamma.
\end{align}
Following some mathematical manipulations defined in \cite{moualeu2018physical} and utilizing identities \cite[Eqs.~(8.4.3.1) and (2.24.1.1)]{Calculationbook02}, $\mathcal{R}_{2}$ and $\mathcal{R}_{3}$ are further derived as
\begin{align}
\nonumber
\mathcal{R}_{2}&=\mathcal{R}_{3}=\int_{0}^{\infty}\gamma^{\frac{\tilde{\alpha_{r}}m_{r}+\Theta_{e}}{\tilde{\alpha_{r}}}}e^{-\delta_{r}\Psi_{T}^{-\tilde{\alpha_{r}}}\sigma^{\tilde{\alpha_{r}}}\gamma}e^{-\delta_{e}\gamma^{\frac{\tilde{\alpha_{e}}}{\tilde{\alpha_{r}}}}}d\left(\gamma^{\frac{1}{\tilde{\alpha_{r}}}}\right)
\\
\nonumber
&=\frac{1}{\tilde{\alpha_{r}}}\int_{0}^{\infty}\gamma^{\Xi_{7}^{II}-1}e^{-\delta_{r}\Psi_{T}^{-\tilde{\alpha_{r}}}\sigma^{\tilde{\alpha_{r}}}\gamma}e^{-\delta_{e}\gamma^{\frac{\tilde{\alpha_{e}}}{\tilde{\alpha_{r}}}}}d\gamma
\\
\nonumber
&=\frac{1}{\tilde{\alpha_{r}}}\int_{0}^{\infty}\gamma^{\Xi_{7}^{II}-1}G_{0,1}^{1,0}\left[\delta_{r}\Psi_{T}^{-\tilde{\alpha_{r}}}\sigma^{\tilde{\alpha_{r}}}\gamma \biggl |
\begin{array}{c}
- \\
0 \\
\end{array}
\right]
\\
\nonumber
&\times G_{0,1}^{1,0}\left[\delta_{e}\gamma^{\frac{\tilde{\alpha_{e}}}{\tilde{\alpha_{r}}}}\biggl |
\begin{array}{c}
- \\
0 \\
\end{array}
\right]d\gamma=\frac{\tilde{\alpha_{e}}^{\Xi_{7}^{II}-\frac{1}{2}}(\delta_{r}\Psi_{T}^{-\tilde{\alpha_{r}}}\sigma^{\tilde{\alpha_{r}}})^{-\Xi_{7}^{II}}}{\tilde{\alpha_{r}}^{\frac{1}{2}}(2\pi)^{\frac{1}{2}(\tilde{\alpha_{e}}+\tilde{\alpha_{r}}-2)}}
\\
&\times G_{\tilde{\alpha_{e}},\tilde{\alpha_{r}}}^{\tilde{\alpha_{r}},\tilde{\alpha_{e}}}\left[\frac{\delta_{e}^{\tilde{\alpha_{r}}}\tilde{\alpha_{r}}^{-\tilde{\alpha_{r}}}}{(\delta_{r}\Psi_{T}^{-\tilde{\alpha_{r}}}\sigma^{\tilde{\alpha_{r}}})^{\tilde{\alpha_{e}}}\tilde{\alpha_{e}}^{-\tilde{\alpha_{e}}}}\biggl |
\begin{array}{c}
\Delta(\tilde{\alpha_{e}}, 1-\Xi_{7}^{II}) \\
0 \\
\end{array}
\right],
\end{align}
where $\Xi_{7}^{II}=\frac{\tilde{\alpha_{r}}m_{r}+\Theta_{e}+\tilde{\alpha_{r}}}{\tilde{\alpha_{r}}}$.

\subsubsection*{Derivation of $\mathcal{R}_{4}$}
By following the similar approach as of $\mathcal{R}_{2}$ and $\mathcal{R}_{3}$, $\mathcal{R}_{4}$ is expressed as
\begin{align}
\nonumber
&\mathcal{R}_{4}=\int_{0}^{\infty}\gamma^{\tilde{\alpha_{r}}(m_{r}+m_{4}+m_{5})+\Theta_{e}}e^{-\delta_{r}\Psi_{T}^{-2\tilde{\alpha_{r}}}\Psi_{Q}^{\tilde{\alpha_{r}}}\sigma^{\tilde{\alpha_{r}}}\gamma^{\tilde{\alpha_{r}}}}e^{-\delta_{e}\gamma^{\tilde{\alpha_{e}}}}d\gamma
\\
\nonumber
&=\int_{0}^{\infty}\gamma^{\frac{\tilde{\alpha_{r}}(m_{r}+m_{4}+m_{5})+\Theta_{e}}{\tilde{\alpha_{r}}}}e^{-\delta_{r}\Psi_{T}^{-2\tilde{\alpha_{r}}}\Psi_{Q}^{\tilde{\alpha_{r}}}\sigma^{\tilde{\alpha_{r}}}\gamma}e^{-\delta_{e}\gamma^{\frac{\tilde{\alpha_{e}}}{\tilde{\alpha_{r}}}}}d\left(\gamma^{\frac{1}{\tilde{\alpha_{r}}}}\right)
\\
\nonumber
&=\frac{1}{\tilde{\alpha_{r}}}\int_{0}^{\infty}\gamma^{\Xi_{8}^{II}-1}e^{-\delta_{r}\Psi_{T}^{-2\tilde{\alpha_{r}}}\Psi_{Q}^{\tilde{\alpha_{r}}}\sigma^{\tilde{\alpha_{r}}}\gamma}e^{-\delta_{e}\gamma^{\frac{\tilde{\alpha_{e}}}{\tilde{\alpha_{r}}}}}d \gamma
\\
\nonumber
&=\frac{1}{\tilde{\alpha_{r}}}\int_{0}^{\infty}\gamma^{\Xi_{8}^{II}-1}G_{0,1}^{1,0}\left[\delta_{r}\Psi_{T}^{-2\tilde{\alpha_{r}}}\Psi_{Q}^{\tilde{\alpha_{r}}}\sigma^{\tilde{\alpha_{r}}}\gamma \biggl |
\begin{array}{c}
- \\
0 \\
\end{array}
\right]
\\
\nonumber
&\times G_{0,1}^{1,0}\left[\delta_{e}\gamma^{\frac{\tilde{\alpha_{e}}}{\tilde{\alpha_{r}}}}\biggl |
\begin{array}{c}
- \\
0 \\
\end{array}
\right]d \gamma=\frac{\tilde{\alpha_{e}}^{\Xi_{8}^{II}-\frac{1}{2}}(\delta_{r}\Psi_{T}^{-2\tilde{\alpha_{r}}}\Psi_{Q}^{\tilde{\alpha_{r}}}\sigma^{\tilde{\alpha_{r}}})^{-\Xi_{8}^{II}}}{\tilde{\alpha_{r}}^{\frac{1}{2}}(2\pi)^{\frac{1}{2}(\tilde{\alpha_{e}}+\tilde{\alpha_{r}}-2)}}
\\
&\times G_{\tilde{\alpha_{e}},\tilde{\alpha_{r}}}^{\tilde{\alpha_{r}},\tilde{\alpha_{e}}}\left[\frac{\delta_{e}^{\tilde{\alpha_{r}}}\tilde{\alpha_{r}}^{-\tilde{\alpha_{r}}}}{(\delta_{r}\Psi_{T}^{-\tilde{2\alpha_{r}}}\Psi_{Q}^{\tilde{\alpha_{r}}}\sigma^{\tilde{\alpha_{r}}})^{\tilde{\alpha_{e}}}\tilde{\alpha_{e}}^{-\tilde{\alpha_{e}}}}\biggl |
\begin{array}{c}
\Delta(\tilde{\alpha_{e}}, 1-\Xi_{8}^{II}) \\
0 \\
\end{array}
\right],
\end{align}
where $\Xi_{8}^{II}=\frac{\tilde{\alpha_{r}}(m_{r}+m_{4}+m_{5})+\Theta_{e}+\tilde{\alpha_{r}}}{\tilde{\alpha_{r}}}$.

\subsubsection*{Derivation of $\mathcal{R}_{5}$}
With some mathematical manipulations and following identities \cite[Eqs.~(8.4.3.1) and (2.24.1.1)]{Calculationbook02}, $\mathcal{R}_{5}$ is derived as
\begin{align}
\nonumber
&\mathcal{R}_{5}=\int_{0}^{\infty}\gamma^{\Theta_{e}}e^{-\delta_{e}\gamma^{\tilde{\alpha_{e}}}}G_{s+1,3s+1}^{3s,1}\left[\frac{V\sigma\gamma}{\mu_{s}}\biggl |
\begin{array}{c}
1,q_{1} \\
q_{2},0 \\
\end{array}
\right]d\gamma
\\
\nonumber
&=\int_{0}^{\infty}\gamma^{\Theta_{e}}G_{0,1}^{1,0}\left[\delta_{e}\gamma^{\tilde{\alpha_{e}}}\biggl |
\begin{array}{c}
- \\
0 \\
\end{array}
\right]G_{s+1,3s+1}^{3s,1}\left[\frac{V\sigma\gamma}{\mu_{s}}\biggl |
\begin{array}{c}
 1,q_{1} \\
 q_{2},0 \\
\end{array}
\right]\ d\gamma
\\
\nonumber
&=\frac{\tilde{\alpha_{e}}^{s(2\Theta_{e}+1)}\left(\frac{V\sigma}{\mu_{s}}\right)^{-(\Theta_{e}+1)}}{(2\pi)^{s(\tilde{\alpha_{e}}-1)}}
\\
&\times G_{\tilde{\alpha_{e}}(3s+1),1+\tilde{\alpha_{e}}(s+1)}^{1+\tilde{\alpha_{e}}, \tilde{3s\alpha_{e}}}\left[\frac{\delta_{e}\tilde{\alpha_{e}}^{2s\tilde{\alpha_{e}}}}{\left(\frac{V\sigma}{\mu_{s}}\right)^{\tilde{\alpha_{e}}}}\Biggl|
\begin{array}{c}
\zeta_{1}, \Delta(\tilde{\alpha_{e}},-\Theta_{e}) \\
0, \zeta_{2}, \zeta_{3} \\
\end{array}
\right].
\end{align}

\subsubsection*{Derivation of $\mathcal{R}_{6}$ and $\mathcal{R}_{7}$}
$\mathcal{R}_{6}$ and $\mathcal{R}_{7}$ are derived by following
\cite[Eq.~(8.4.3.1)]{Calculationbook02} as
\begin{align}
\nonumber
\mathcal{R}_{6}=\mathcal{R}_{7}&=\int_{0}^{\infty}\gamma^{\tilde{\alpha_{r}}m_{r}+\Theta_{e}}e^{-\delta_{r}\Psi_{T}^{-\tilde{\alpha_{r}}}\sigma^{\tilde{\alpha_{r}}}\gamma^{\tilde{\alpha_{r}}}}e^{-\delta_{e}\gamma^{\tilde{\alpha_{e}}}}
\\
\nonumber
&\times G_{s+1,3s+1}^{3s,1}\left[\frac{V\sigma\gamma}{\mu_{s}}\biggl |
\begin{array}{c}
1,q_{1} \\
q_{2},0 \\
\end{array}
\right] d\gamma
\\
\nonumber
&=\int_{0}^{\infty}\gamma^{\tilde{\alpha_{r}}m_{r}+\Theta_{e}}G_{0,1}^{1,0}\left[\delta_{r}\Psi_{T}^{-\tilde{\alpha_{r}}}\sigma^{\tilde{\alpha_{r}}}\gamma^{\tilde{\alpha_{r}}} \biggl |
\begin{array}{c}
- \\
0 \\
\end{array}
\right]
\\
&\times G_{0,1}^{1,0}\left[\delta_{e}\gamma^{\tilde{\alpha_{e}}}\biggl |
\begin{array}{c}
- \\
0 \\
\end{array}
\right] G_{s+1,3s+1}^{3s,1}\left[\frac{V\sigma\gamma}{\mu_{s}}\biggl |
\begin{array}{c}
1,q_{1} \\
q_{2},0 \\
\end{array}
\right] d\gamma.
\end{align}
Now, for tractable analysis, transforming Meijer's $G$ functions into Fox's $H$ functions via utilizing \cite[Eqs.~(6.2.3) and (6.2.8)]{springer1979algebra} and performing integration by means of \cite[Eq.~(2.3)]{mittal1972integral} and \cite[Eq.~(3)]{lei2017secrecyal}, $\mathcal{R}_{6}$ and $\mathcal{R}_{7}$ are obtained as
\begin{align}
\nonumber
&\mathcal{R}_{6}=\mathcal{R}_{7}=\int_{0}^{\infty}\gamma^{\tilde{\alpha_{r}}m_{r}+\Theta_{e}}H_{s+1,3s+1}^{3s,1}\left[\frac{V\sigma\gamma}{\mu_{s}}\biggl |
\begin{array}{c}
(1,1), (q_{1},1) \\
(q_{2},1), (0,1) \\
\end{array}
\right]
\\
\nonumber
&\times H_{0,1}^{1,0}\left[\Xi_{10}^{II}\gamma^{\tilde{\alpha_{r}}} \biggl |
\begin{array}{c}
- \\
(0,1) \\
\end{array}
\right] H_{0,1}^{1,0}\left[\delta_{e}\gamma^{\tilde{\alpha_{e}}}\biggl |
\begin{array}{c}
- \\
(0,1) \\
\end{array}
\right]  d\gamma=\frac{\delta_{e}^{-\frac{\Xi_{9}^{II}}{\tilde{\alpha_{e}}}}}{\tilde{\alpha_{e}}}
\\
&\times H_{1,0:0,1:s+1,3s+1}^{1,0:1,0:3s,1}\biggl[\begin{array}{c}
J_{1}\\
J_{2}\\
\end{array}\biggl | \begin{array}{c}
-\\
(0,1)\\
\end{array}\biggl | \begin{array}{c}
(1,1), (q_{1},1)\\
(q_{2},1), (0,1)\\
\end{array}\biggl | J_{3},J_{4}\biggl],
\end{align}
where $\Xi_{9}^{II}=\tilde{\alpha_{r}}m_{r}+\Theta_{e}+1$, $\Xi_{10}^{II}=\delta_{r}\Psi_{T}^{-\tilde{\alpha_{r}}}\sigma^{\tilde{\alpha_{r}}}$, $J_{1}=\left(1-\frac{\Xi_{9}^{II}}{\tilde{\alpha_{e}}};\frac{\tilde{\alpha_{r}}}{\tilde{\alpha_{e}}},\frac{1}{\tilde{\alpha_{e}}}\right)$, $J_{2}=(1;-)$, $J_{3}=\Xi_{10}^{II}\delta_{e}^{-\frac{\tilde{\alpha_{r}}}{\tilde{\alpha_{e}}}}$, $J_{4}=V\sigma\mu_{s}^{-1}\delta_{e}^{-\frac{1}{\tilde{\alpha_{e}}}}$, and $H_{p,q}^{m,n}[\cdot]$ is the Fox's $H$ function introduced in \cite[Eq.~(1.2)]{mathai2009h}, and $H_{c_{1},d_{1}:c_{2},d_{2}:c_{3},d_{3}}^{x_{1},y_{1}:x_{2},y_{2}:x_{3},y_{3}}[\cdot]$ is the extended generalized bivariate Fox’s $H$ function (EGBFHF) as explained in \cite[Eq.~(2.57)]{mathai2009h}.

\subsubsection*{Derivation of $\mathcal{R}_{8}$}

$\mathcal{R}_{8}$ is derived by following the similar identities from $\mathcal{R}_{6}$ and $\mathcal{R}_{7}$ as
\begin{align}
\nonumber
&\mathcal{R}_{8}=\int_{0}^{\infty}\gamma^{\tilde{\alpha_{r}}(m_{r}+m_{4}+m_{5})+\Theta_{e}}e^{-\delta_{r}\Psi_{T}^{-2\tilde{\alpha_{r}}}\Psi_{Q}^{\tilde{\alpha_{r}}}\sigma^{\tilde{\alpha_{r}}}\gamma^{\tilde{\alpha_{r}}}}e^{-\delta_{e}\gamma^{\tilde{\alpha_{e}}}}
\\
\nonumber
&\times G_{s+1,3s+1}^{3s,1}\left[\frac{V\sigma\gamma}{\mu_{s}}\biggl |
\begin{array}{c}
1,q_{1} \\
q_{2},0 \\
\end{array}
\right] d\gamma=\int_{0}^{\infty}\gamma^{\tilde{\alpha_{r}}(m_{r}+m_{4}+m_{5})+\Theta_{e}}
\\
\nonumber
&\times G_{0,1}^{1,0}\left[\Xi_{12}^{II}\gamma^{\tilde{\alpha_{r}}} \biggl |
\begin{array}{c}
- \\
0 \\
\end{array}
\right]\,G_{0,1}^{1,0}\left[\delta_{e}\gamma^{\tilde{\alpha_{e}}}\biggl |
\begin{array}{c}
- \\
0 \\
\end{array}
\right]
\\
\nonumber
&\times G_{s+1,3s+1}^{3s,1}\left[\frac{V\sigma\gamma}{\mu_{s}}\biggl |
\begin{array}{c}
1,q_{1} \\
q_{2},0 \\
\end{array}
\right] d\gamma=\tilde{\alpha_{e}}^{-1}\delta_{e}^{-\frac{\Xi_{11}^{II}}{\tilde{\alpha_{e}}}}
\\
&\times H_{1,0:0,1:s+1,3s+1}^{1,0:1,0:3s,1}\biggl[\begin{array}{c}
J_{5}\\
J_{2}\\
\end{array}\biggl | \begin{array}{c}
-\\
(0,1)\\
\end{array}\biggl | \begin{array}{c}
(1,1), (q_{1},1)\\
(q_{2},1), (0,1)\\
\end{array}\biggl | J_{6},J_{4}\biggl],
\end{align}
where $\Xi_{11}^{II}=\tilde{\alpha_{r}}(m_{r}+m_{4}+m_{5})+\Theta_{e}+1$, $\Xi_{12}^{II}=\delta_{r}\Psi_{T}^{-2\tilde{\alpha_{r}}}\Psi_{Q}^{\tilde{\alpha_{r}}}\sigma^{\tilde{\alpha_{r}}}$, $J_{5}=(1-\frac{\Xi_{11}^{II}}{\tilde{\alpha_{e}}};\frac{\tilde{\alpha_{r}}}{\tilde{\alpha_{e}}},\frac{1}{\tilde{\alpha_{e}}})$, and $J_{6}=\Xi_{12}^{II}\,\delta_{e}^{-\frac{\tilde{\alpha_{r}}}{\tilde{\alpha_{e}}}}$.

\subsection{Strictly Positive Secrecy Capacity Analysis}

In order to ensure secrecy in a wiretap prone model, the probability of SPSC acts as one of the crucial performance metric that ensures uninterrupted data stream conveyance only when secrecy behaviour of the system remains up to scratch ($\gamma_{f,j}>\gamma_{e,j}$). Mathematically, the analytical expression of probability of SPSC can be easily deducted from the SOP that is given as \cite[Eq.~(25)] {islam2020secrecy}
\setcounter{eqnback}{\value{equation}}
\setcounter{equation}{47}
\begin{align}
\nonumber
SPSC&=\Pr\left\{C_{s}>0\right\}
\\
\nonumber
&=1-\Pr\left\{C_{s} \leq 0 \right\}.
\\
\label{eqn:spscs1}
\text{Hence, } SPSC^{I}&=1- \text{SOP}_{L}^{I} |_{\Upsilon_{e}=0}, \text{   (Scenario-I)}
\\
\label{eqn:spscs2}
\text{and } SPSC^{II}&=1- \text{SOP}_{L}^{II} |_{\Upsilon_{e}=0}. \text{   (Scenario-II)}
\end{align}

\subsection{Effective Secrecy Throughput Analysis}

EST is another secrecy performance metric that ensures achievable secure average throughput measurements. When the reliability issue is considered equally important along with the security issue then this parameter is considered. Mathematically, EST is formulated as the product of target secrecy rate and the probability of successful transmission that is given as \cite[Eq.~(5)]{lei2020secure}
\begin{align}
\nonumber
EST&= \Upsilon_{e}(1-SOP).
\\
\label{eqn:ests1}
\text{Hence, } EST^{I} &=\Upsilon_{e}(1-SOP_{L}^{I}), \text{ (Scenario-I)}
\\
\label{eqn:ests2}
\text{and }EST^{II} &=\Upsilon_{e}(1-SOP_{L}^{II}). \text{ (Scenario-II)}
\end{align}

\section{Numerical Results}
\label{results}

In this section, the impacts of all the system parameters (e.g., $\alpha_{i}$, $\mu_{i}$, $\Phi_{i}$, $\Psi_{Q}$, $\Psi_{T}$, $P_{o}$, $\alpha_{o}$, $\beta_{o}$, $\epsilon$, and $s$) on the secrecy behaviour of the proposed scenarios are illustrated graphically utilizing the expressions in \eqref{eqn:sopf1}, \eqref{eqn:sopf}, \eqref{eqn:spscs1}, \eqref{eqn:spscs2}, \eqref{eqn:ests1}, and \eqref{eqn:ests2}. Note that the univariate Meijer's $G$ and Fox's $H$ functions can be easily computed utilizing standard Mathematica packages whereas the EGBFHF is computed utilizing \cite[Table~I]{lei2017secrecyal}. Moreover, all the mathematical expressions are also validated via MC simulations averaging $10^{6}$ random samples of $\alpha-\mu$ and M\'alaga random variables. It can clearly be seen that MC simulation results are in a good agreement with the analytical results that strongly corroborate the analysis of this paper. All the numerical results are demonstrated assuming ($\alpha_{o}$, $\beta_{o}$) $=$ ($2.296, 2$) for strong, ($\alpha_{o}$, $\beta_{o}$) $=$ ($4.2, 3$) for moderate, and ($\alpha_{o}$, $\beta_{o}$) $=$ ($8, 4$) for weak turbulences, $r=1$ for HD and $r=2$ for IM/DD techniques, and $\Upsilon_{e}=0.05$ bits/sec/Hz.

\begin{figure}[!h]
\vspace{-10mm}
    \centerline{\includegraphics[width=0.75\textwidth,angle =0]{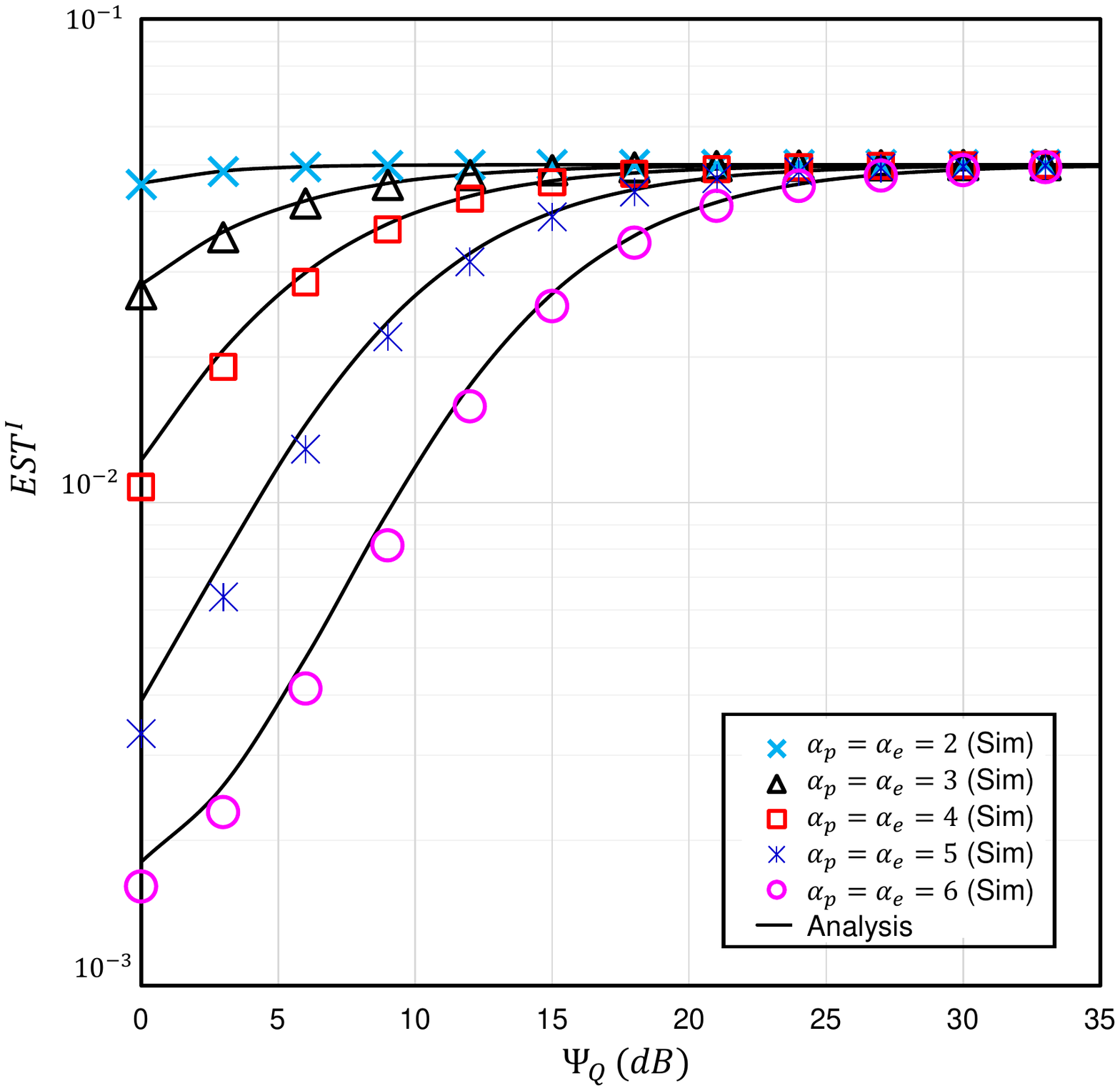}}
        \vspace{-10mm }
    \caption{
        The EST versus $\Psi_{Q}$ for selected values of $\alpha_{p}$ and $\alpha_{e}$ with $\alpha_{o}=2.296$, $\beta_{o}=2$, $\Omega_{o}=1$, $g=2$, $s=2$, $\epsilon=1$, $P_{o}=0.5$, $\alpha_{r}=2$, $\mu_{r}=\mu_{p}=\mu_{e}=2$, $\Phi_{r}=15dB$, $\Phi_{p}=10dB$, $\Phi_{e}=\mu_{s}-5dB$, and $\Upsilon_{e}=0.05$ bits/sec/Hz.
    }
    \label{g11}
\end{figure}

\begin{figure}[!h]
\vspace{-10mm}
    \centerline{\includegraphics[width=0.75\textwidth,angle =0]{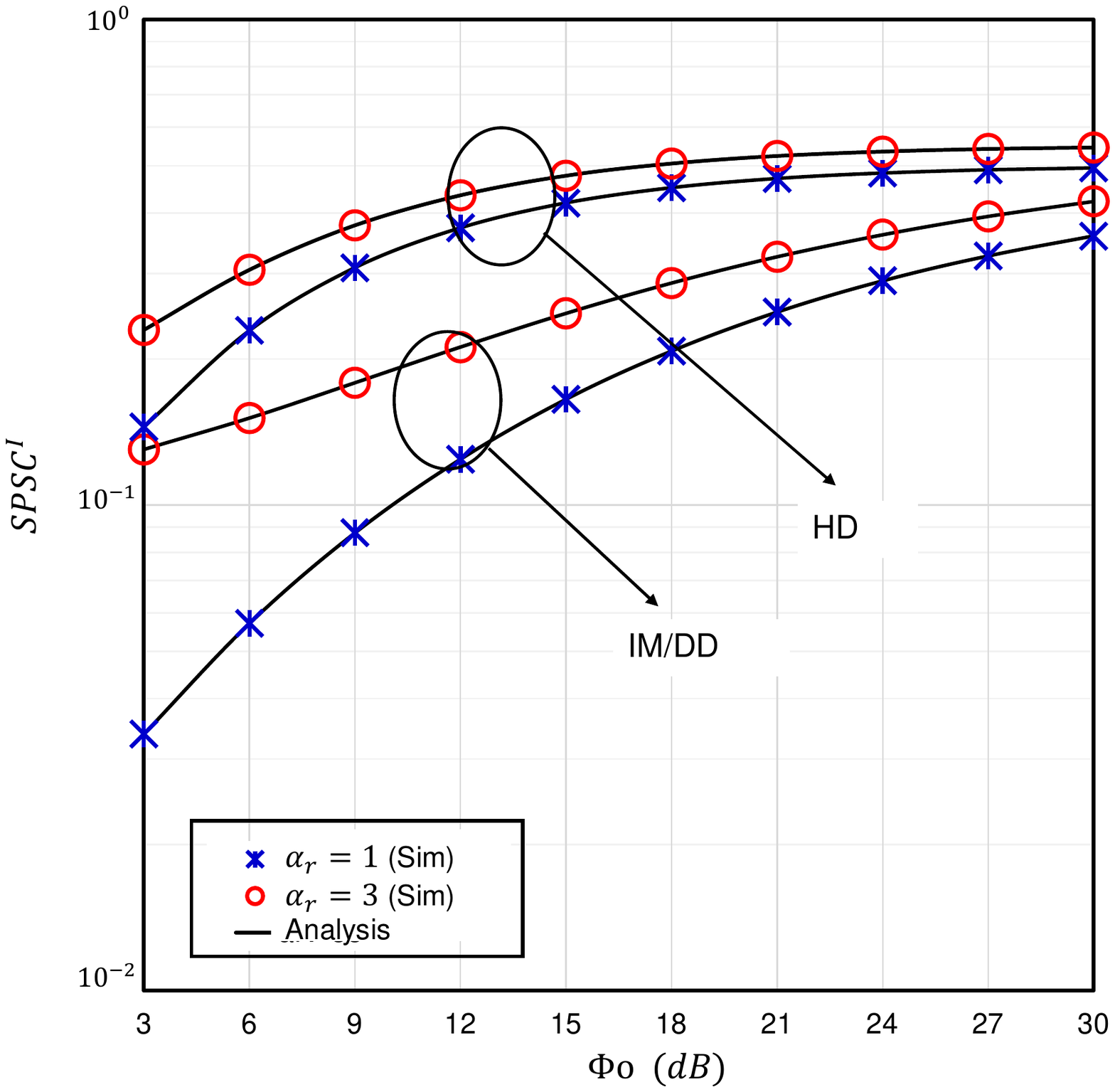}}
        \vspace{-10mm }
    \caption{
         The SPSC versus $\Phi_{o}$ for selected values of $\alpha_{r}$ and $s$ with $\alpha_{o}=2.296$,  $\beta_{o}=2$, $\Omega_{o}=1$, $g=2$, $\epsilon=1$, $P_{o}=0.5$, $\alpha_{p}=\alpha_{e}=5$, $\mu_{r}=\mu_{p}=\mu_{e}=6$, $\Phi_{r}=\Phi_{p}=15dB$, $\Phi_{e}=0 dB$, and $\Psi_{Q}=15 dB$.
    }
    \label{g12}
\end{figure}

The impacts of $\alpha_{i}$ on the EST and SPSC are demonstrated in Figs. \ref{g11} and \ref{g12} for Scenario-I.

\begin{figure}[!h]
\vspace{-30mm}
    \centerline{\includegraphics[width=0.55\textwidth,angle =0]{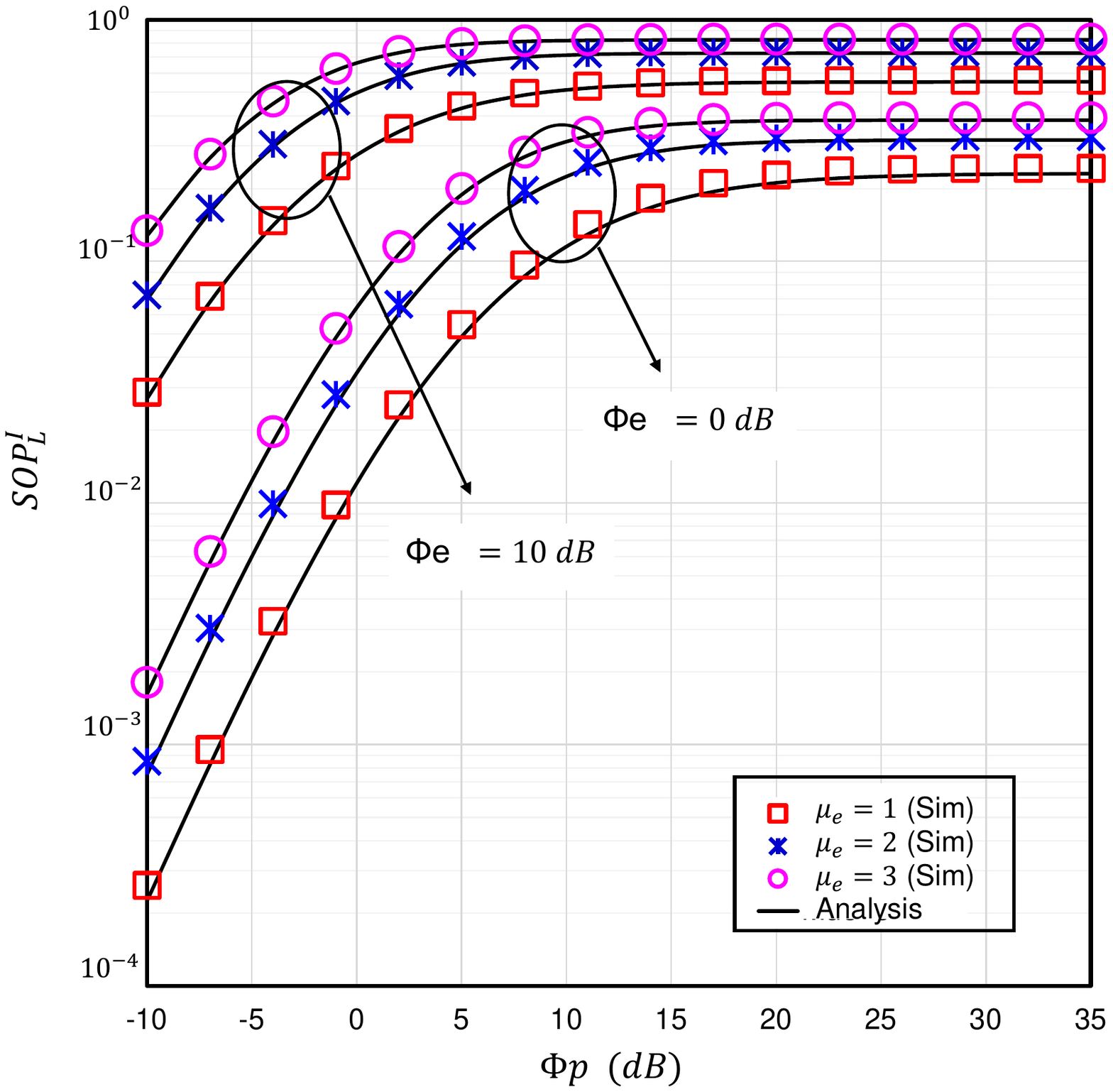}}
        \vspace{-20mm }
    \caption{The SOP versus $\Phi_{p}$ for selected values of $\mu_{e}$ and $\Phi_{e}$ with $\alpha_{o}=2.296$, $\beta_{o}=2$, $\Omega_{o}=1$, $g=2$, $s=1$, $\epsilon=1$, $\alpha_{r}=\alpha_{p}=\alpha_{e}=2$, $\mu_{r}=\mu_{p}=2$, $P_{o}=0.1$, $\Phi_{r}=15dB$, $\mu_{r}=12dB$, $\Psi_{Q}=-5dB$, and $\Upsilon_{e}=0.05$ bits/sec/Hz.
    }
    \label{g21}
\end{figure}
It is observed that EST gradually increases with $\Psi_{Q}$. Similar observations are also seen in \cite{7422832}. Again, since with the increase in non-linearity parameter, the communication link becomes better \cite{juel2021secrecy} (i.e. the link will undergo less amount of fading with higher $\alpha_{i}$), the EST becomes better for lower $\alpha_{p}$ and $\alpha_{e}$. On the other hand, as expected, the SPSC is enhanced as $\Phi_{o}$ and $\alpha_{r}$ increase from a lower to a higher value.

\begin{figure}[!h]
\vspace{-10mm}
    \centerline{\includegraphics[width=0.75\textwidth,angle =0]{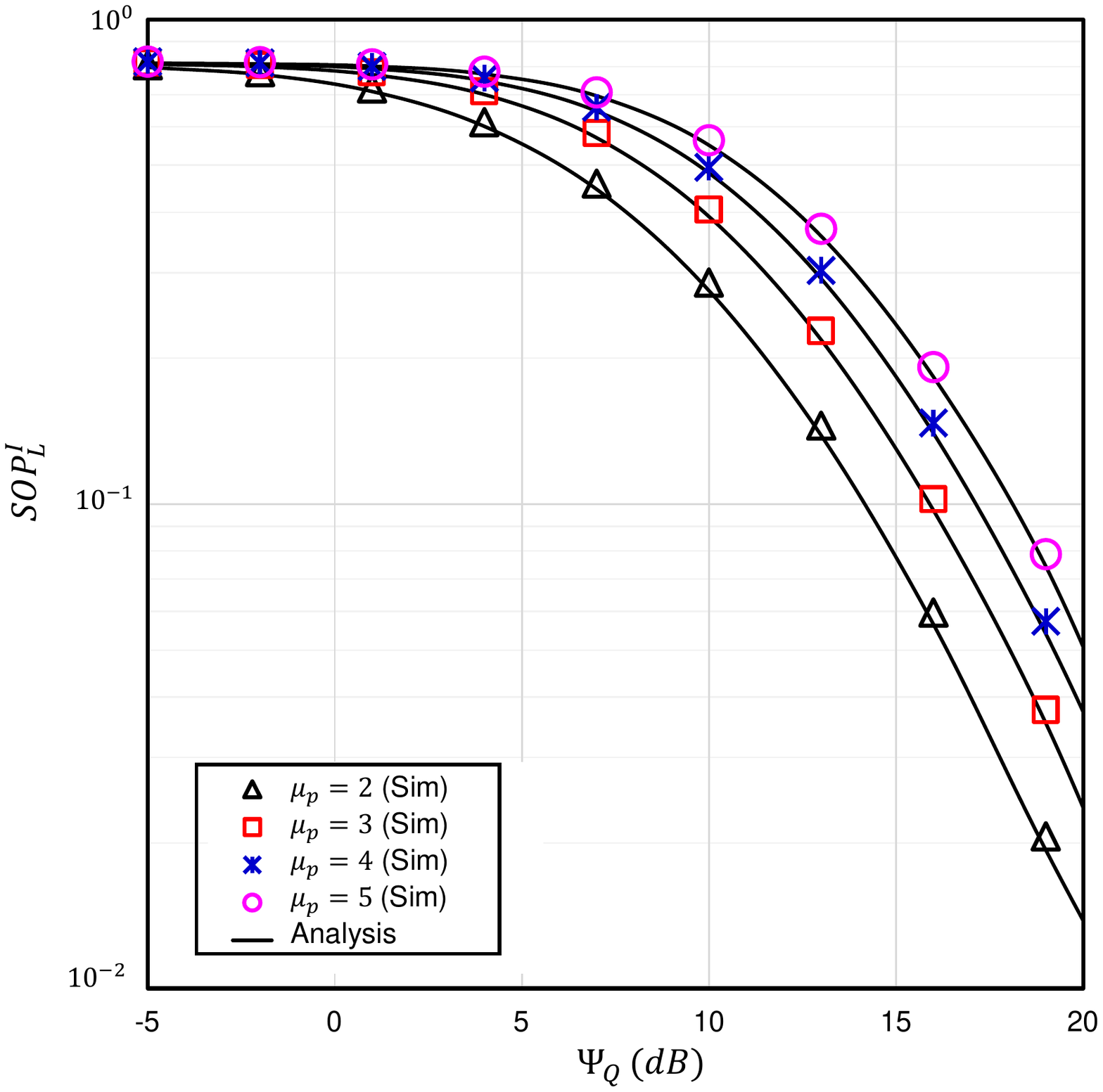}}
        \vspace{-10mm }
    \caption{
         The SOP versus $\Psi_{Q}$ for selected values of $\mu_{p}$ with $\alpha_{o}=2.296$, $\beta_{o}=2$, $\Omega_{o}=1$, $g=2$, $\epsilon=1$, $s=1$, $\alpha_{r}=\alpha_{p}=\alpha_{e}=2$, $\mu_{r}=\mu_{e}=2$, $P_{o}=0.1$, $\Phi_{p}=10dB$, $\Phi_{r}=15dB$ $\Phi_{o}=10dB$, $\Phi_{e}=10dB$, and $\Upsilon_{e}=0.05$ bits/sec/Hz.
    }
    \label{g22}
\end{figure}

\begin{figure}[!h]
\vspace{-10mm}
    \centerline{\includegraphics[width=0.75\textwidth,angle =0]{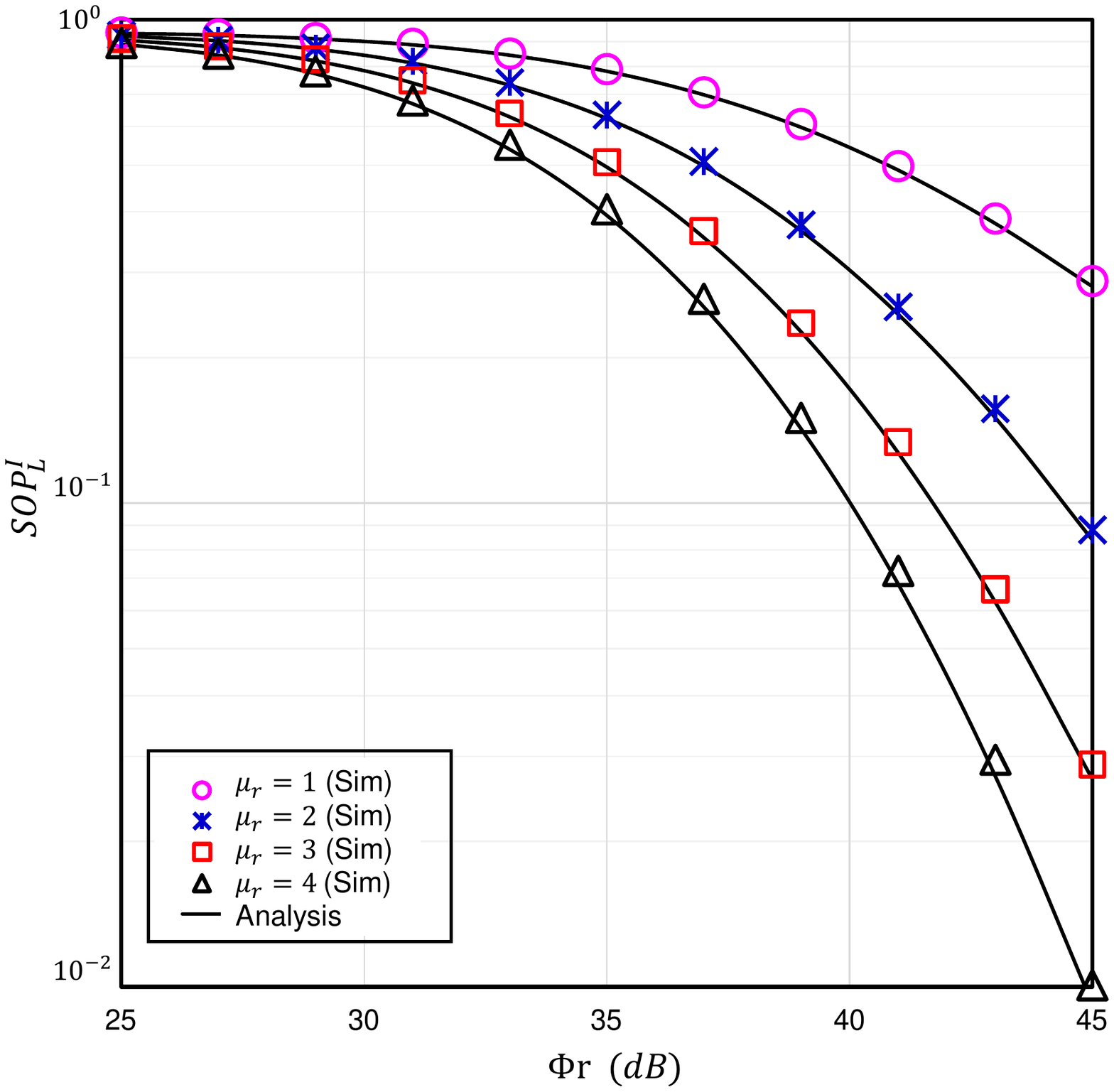}}
        \vspace{-10mm }
    \caption{
         The SOP versus $\Phi_{r}$ for selected values of $\mu_{r}$ with $\alpha_{o}=2.296$, $\beta_{o}=2$, $\Omega_{o}=1$, $g=2$, $\epsilon_{o}=1$, $s=1$, $\alpha_{r}=\alpha_{p}=\alpha_{e}=2$, $\mu_{p}=\mu_{e}=2$, $P_{o}=0.1$, $\Phi_{p}=15dB$, $\Phi_{o}=10dB$, $\Phi_{e}=15dB$, $\Psi_{Q}=-5dB$, and $\Upsilon_{e}=0.05$ bits/sec/Hz.
    }
    \label{g23}
\end{figure}

The SOP is plotted against $\Phi_{p}, \Psi_{Q}$, and $\Phi_{r}$ considering Scenario-I in Figs. \ref{g21}, \ref{g22}, and \ref{g23}, respectively, with a view to observe the impacts of the number of multipath clusters.

\begin{figure}[!h]
\vspace{-10mm}
    \centerline{\includegraphics[width=0.75\textwidth,angle =0]{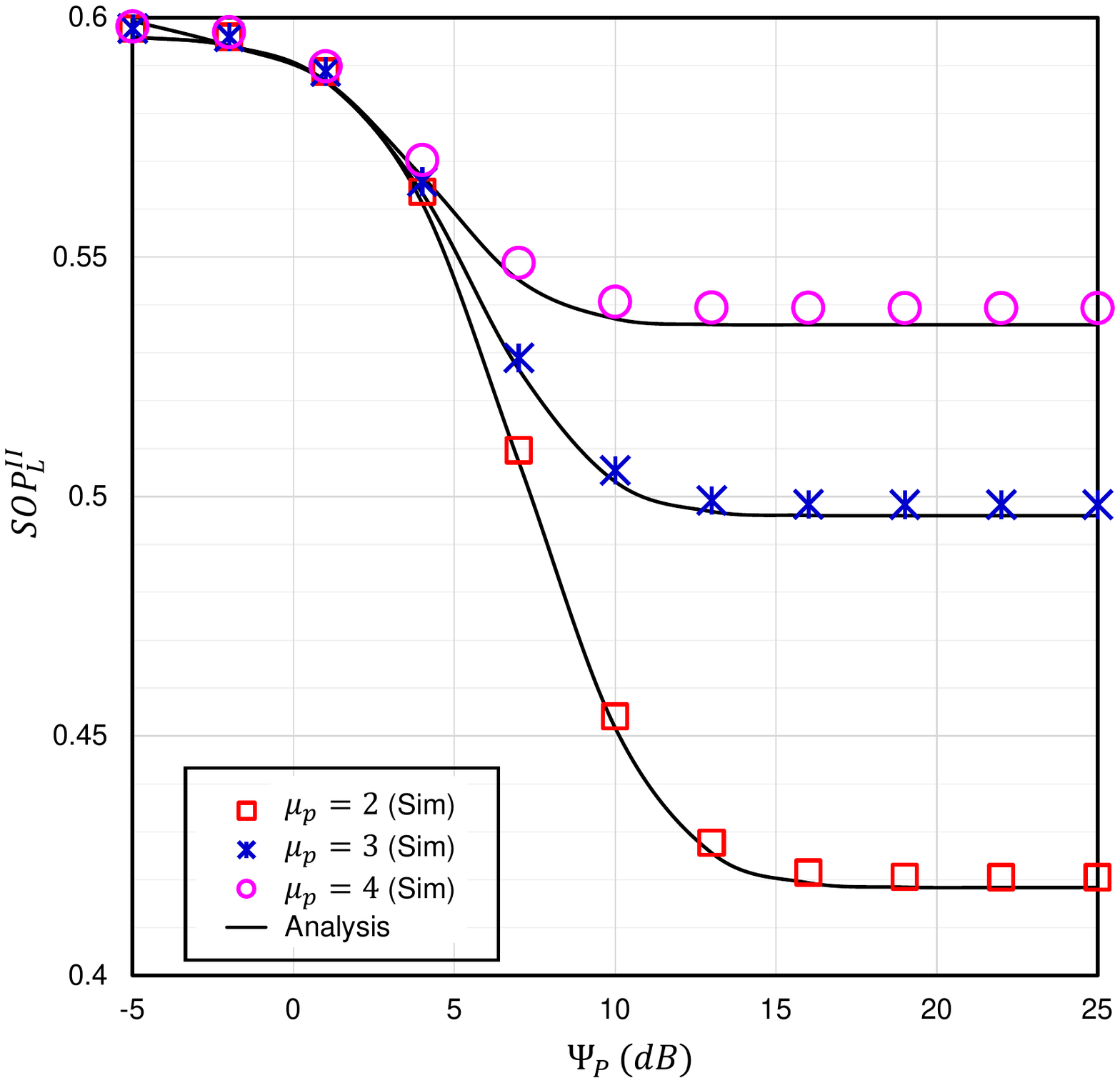}}
        \vspace{-10mm }
    \caption{
         The SOP versus $\Psi_{T}$ for selected values of $\mu_{p}$ with $\alpha_{o}=2.296$, $\beta_{o}=2$, $\Omega_{o}=1$, $g=2$, $\epsilon=6.7$, $P_{o}=0.2$, $s=1$, $\alpha_{r}=\alpha_{p}=\alpha_{e}=2$, $\mu_{r}=\mu_{e}=2$, $\Phi_{p}=-5dB$, $\mu_{s}=10dB$, $\Phi_{r}=-5dB$, $\Psi_{Q}=5dB$, $\Phi_{e}=5dB$, and $\Upsilon_{e}=0.05$ bits/sec/Hz.
    }
    \label{g24}
\end{figure}

It can clearly be noted that the SOP performance improves with $\Psi_{Q}$ and $\Phi_{r}$, and degrades with $\Phi_{p}$ as testified in \cite{erdogan2019error}. This is as expected because the larger $\Psi_{Q}$ indicates the higher transmitting power at $S$. On the other hand, $\Phi_{r}$ improves the $S-R$ RF sub-link whereas $\Phi_{p}$ strengths the $S-P$ link. It is obvious that an increase in $\mu_{i}$ will increase the channel diversity thereby mitigating the channel fading. Hence, the SOP decreases with $\mu_{r}$, and increases with $\mu_{p}$ and $\mu_{e}$. Similar results were also demonstrated in \cite{juel2021secrecy} that corroborate the outcomes in this work. In Fig. \ref{g24}, impact of $\mu_{p}$ is demonstrated on SOP considering Scenario-II. Similar to Scenario-I, in this particular case, it is also observed that the SOP drastically degrades with $\mu_{p}$.

\begin{figure}[!h]
\vspace{-12mm}
    \centerline{\includegraphics[width=0.75\textwidth,angle =0]{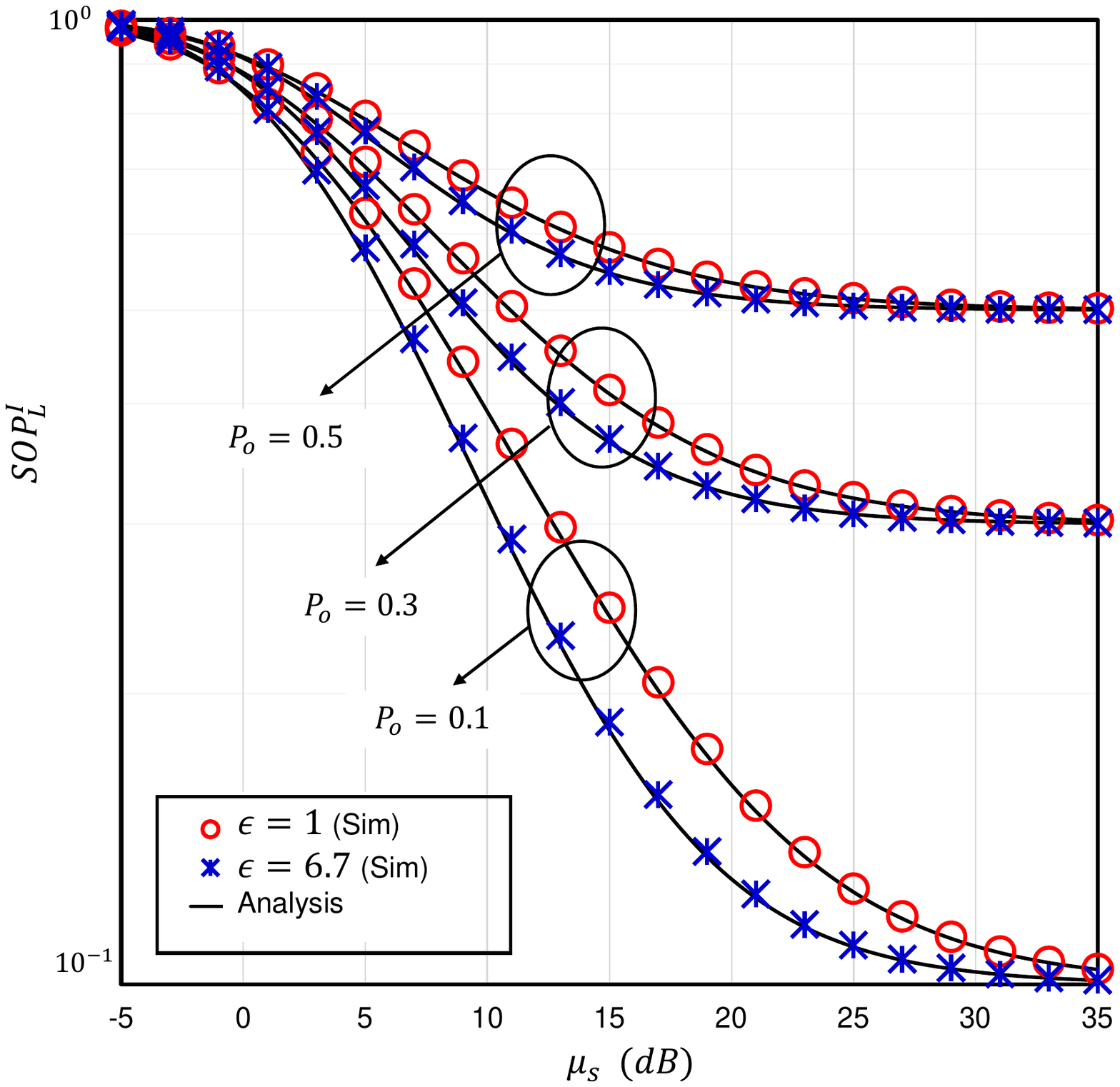}}
        \vspace{-10mm }
    \caption{
         The SOP versus $\mu_{s}$ for selected values of $P_{o}$ and $\epsilon$ with $\alpha_{o}=2.296$, $\beta_{o}=2$, $\Omega_{o}=1$, $g=2$, $s=1$, $\alpha_{r}=\alpha_{p}=\alpha_{e}=2$, $\mu_{r}=\mu_{p}=\mu_{e}=6$, $\Phi_{r}=\Phi_{p}=15dB$, $\Phi_{e}= -5dB$, $\Psi_{Q}=-10dB$, and $\Upsilon_{e}=0.05$ bits/sec/Hz.
    }
    \label{g31}
\end{figure}

In Fig. \ref{g31}, the probability of link blockage is observed via plotting SOP against average SNR of FSO link under various levels of pointing error.
It is observed that probability of link blockage impose significant impact on the SOP performance, that is, the SOP degrades with the increase in $P_{o}$ (increased $P_{o}$ signifies a stronger link blockage). It is obvious because similar to the observations in \cite{djordjevic2016outage}, the atmospheric turbulence degrades with the increase in link blockage, which yields to a degraded secrecy performance.

\begin{figure}[!h]
\vspace{-10mm}
    \centerline{\includegraphics[width=0.75\textwidth,angle =0]{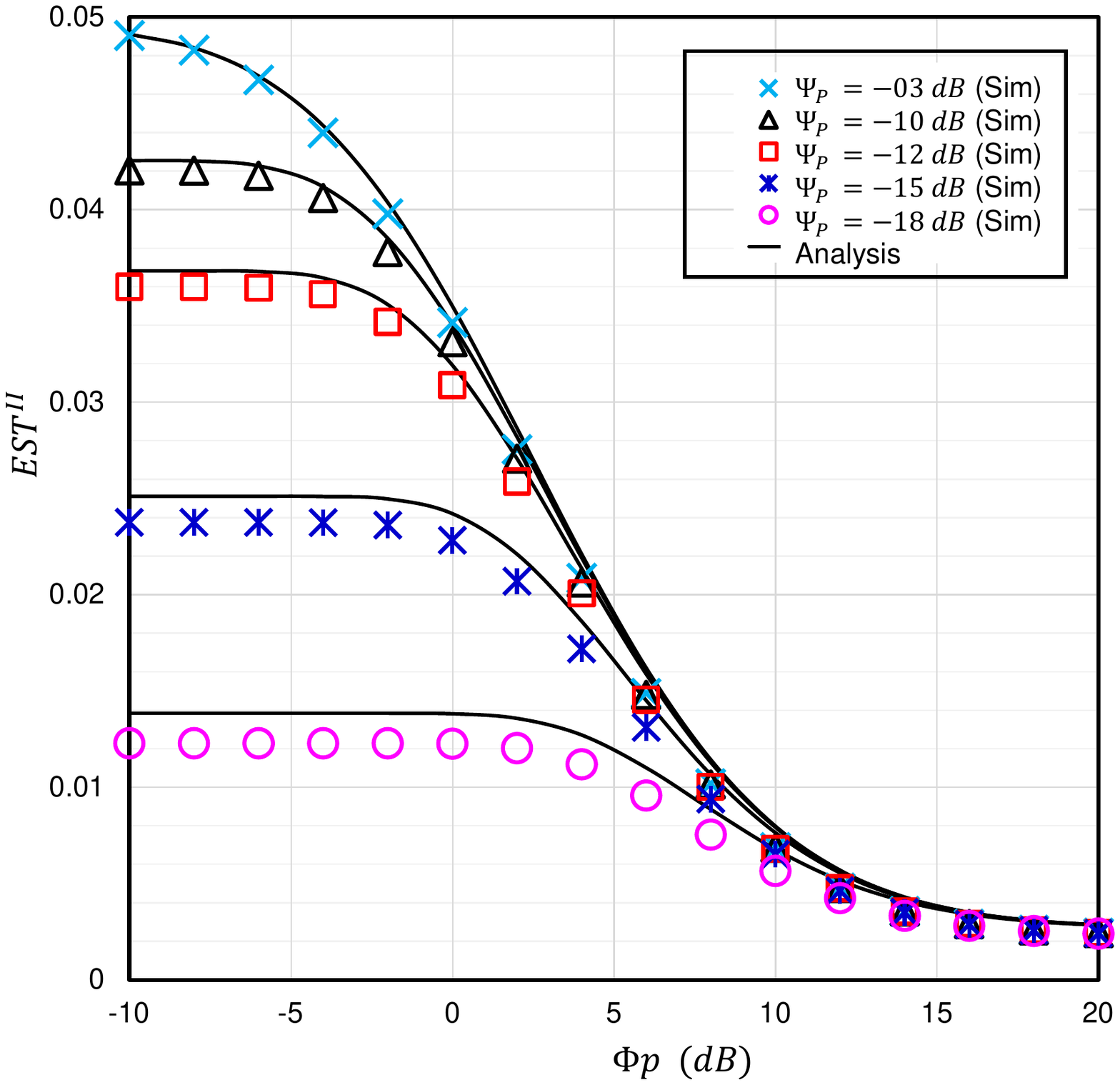}}
        \vspace{-10mm }
    \caption{
         The EST versus $\Phi_{p}$ for selected values of $\Psi_{T}$ with $\alpha_{o}=2.296$, $\beta_{o}=2$, $\Omega_{o}=1$, $g=2$, $s=2$, $\epsilon=6.7$, $P_{o}=0.1$, $\alpha_{r}=\alpha_{p}=\alpha_{e}=2$, $\mu_{r}=\mu_{p}=\mu_{e}=2$, $\Phi_{o}=-5dB$, $\Phi_{r}=10 dB$, $\Psi_{Q}=-10 dB$, $\Phi_{e}=-5 dB$, and $\Upsilon_{e}=0.05$ bits/sec/Hz.
    }
    \label{g41}
\end{figure}

Figure \ref{g41} depicts the impact of $\Psi_{T}$ on the EST performance for Scenario-II.
\begin{figure}[!h]
\vspace{-10mm}
    \centerline{\includegraphics[width=0.75\textwidth,angle =0]{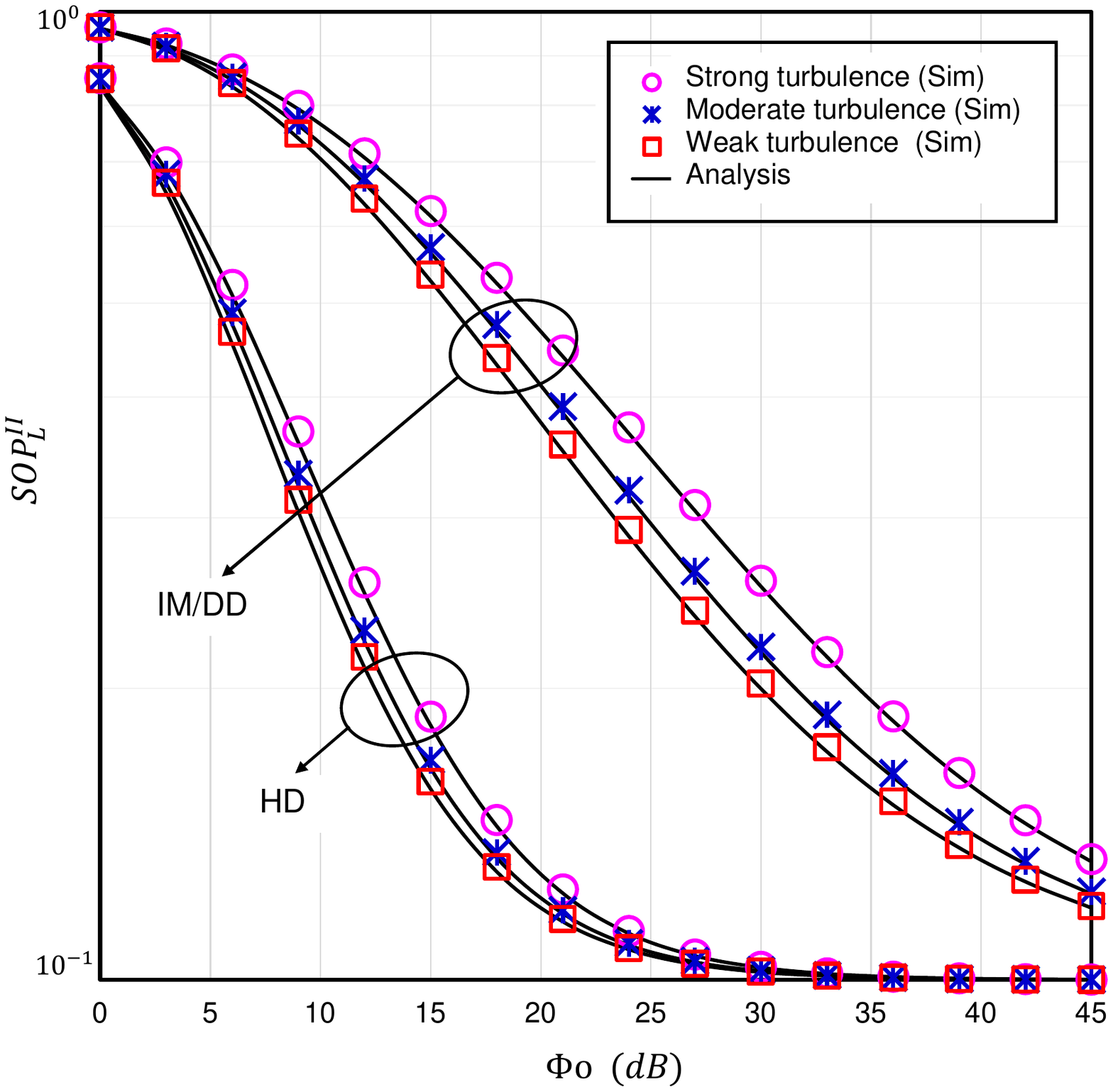}}
        \vspace{-10mm }
    \caption{
          The SOP versus $\Phi_{o}$ for selected values of $\alpha_{o}$, $\beta_{o}$, and $s$ with $\Omega_{o}=1$, $g=2$, $\epsilon=6.7$, $P_{o}=0.1$, $\alpha_{r}=\alpha_{p}=\alpha_{e}=2$, $\mu_{r}=\mu_{p}=\mu_{e}=6$, $\Phi_{r}=\Phi_{p}=15dB$, $\Phi_{e}=-5dB$, $\Psi_{T}=\Psi_{Q}=-10dB$, and $\Upsilon_{e}=0.05$ bits/sec/Hz.
    }
    \label{g51}
\end{figure}
It can be noted that an increased $\Psi_{T}$ is beneficial for EST performance, since for that case the increased transmit power yields a better received SNR at the destination. After $15dB$, an error floor is observed, as expected. This occurs as the transmit power at $S$ is limited by a threshold value.

\begin{figure}[!h]
\vspace{-10mm}
\centerline{\includegraphics[width=0.75\textwidth,angle =0]{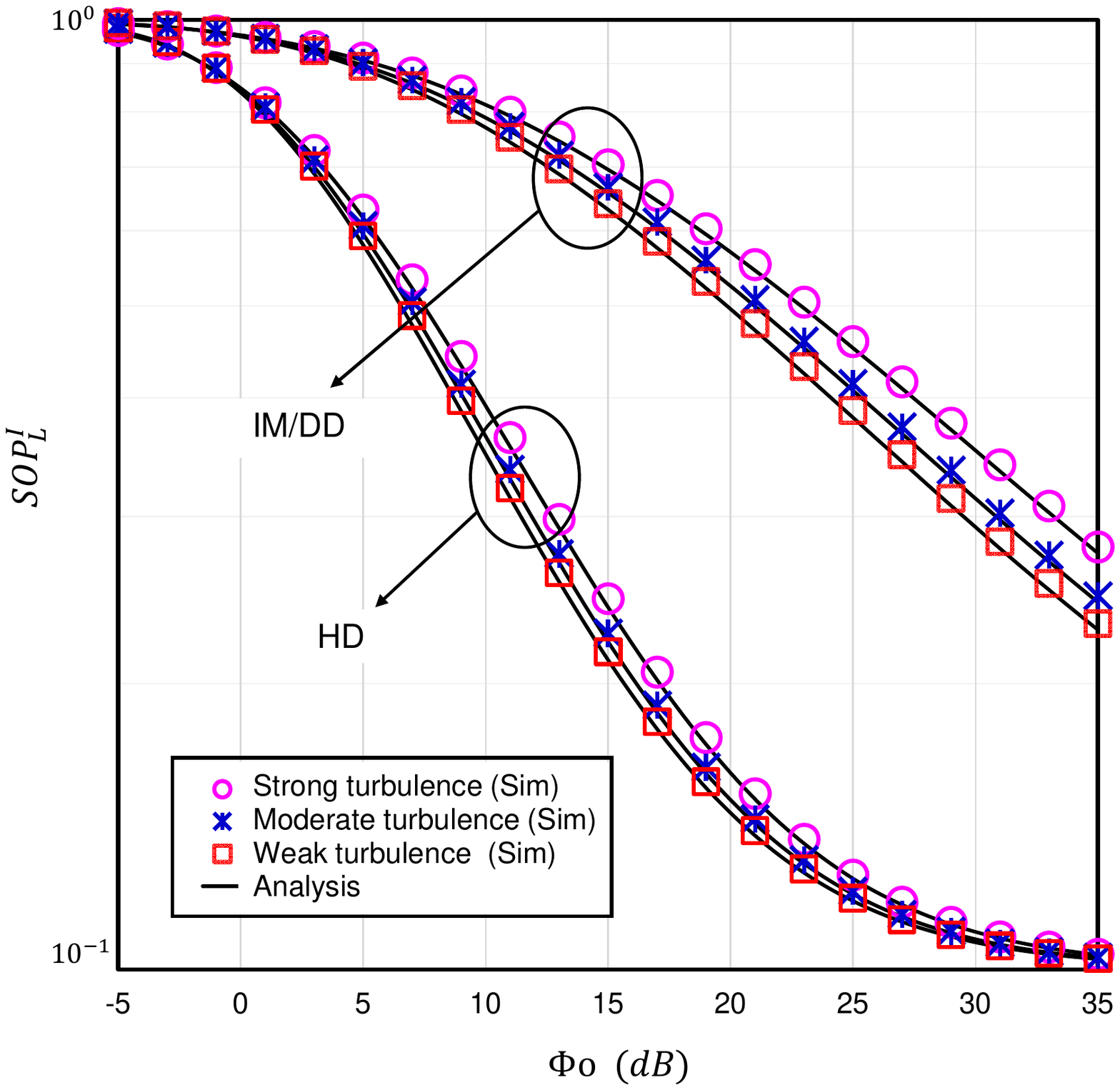}}
        \vspace{-10mm }
    \caption{
          The SOP versus $\Phi_{o}$ for selected values of $\alpha_{o}$, $\beta_{o}$ and $s$ with $\Omega_{o}=1$, $g=2$, $\epsilon=1$, $P_{o}=0.1$, $\alpha_{r}=\alpha_{p}=\alpha_{e}=2$, $\mu_{r}=\mu_{p}=\mu_{e}=6$,$\Phi_{r}=\Phi_{p}=15dB$, $\Phi_{e}=-5dB$, $\Psi_{Q}=-10dB$, and $\Upsilon_{e}=0.05$ bits/sec/Hz.
    }
    \label{g52}
\end{figure}

The FSO link performance is significantly affected by the atmospheric turbulence. Based on the values of $\alpha_{o}$ and $\beta_{o}$, three turbulence conditions (e.g. strong, moderate, and weak) are considered in this work. Figures \ref{g51} and \ref{g52} demonstrate how the secrecy performance is affected with various turbulence conditions under Scenarios- I \& II.

\begin{figure}[!h]
\vspace{-10mm}
    \centerline{\includegraphics[width=0.75\textwidth,angle =0]{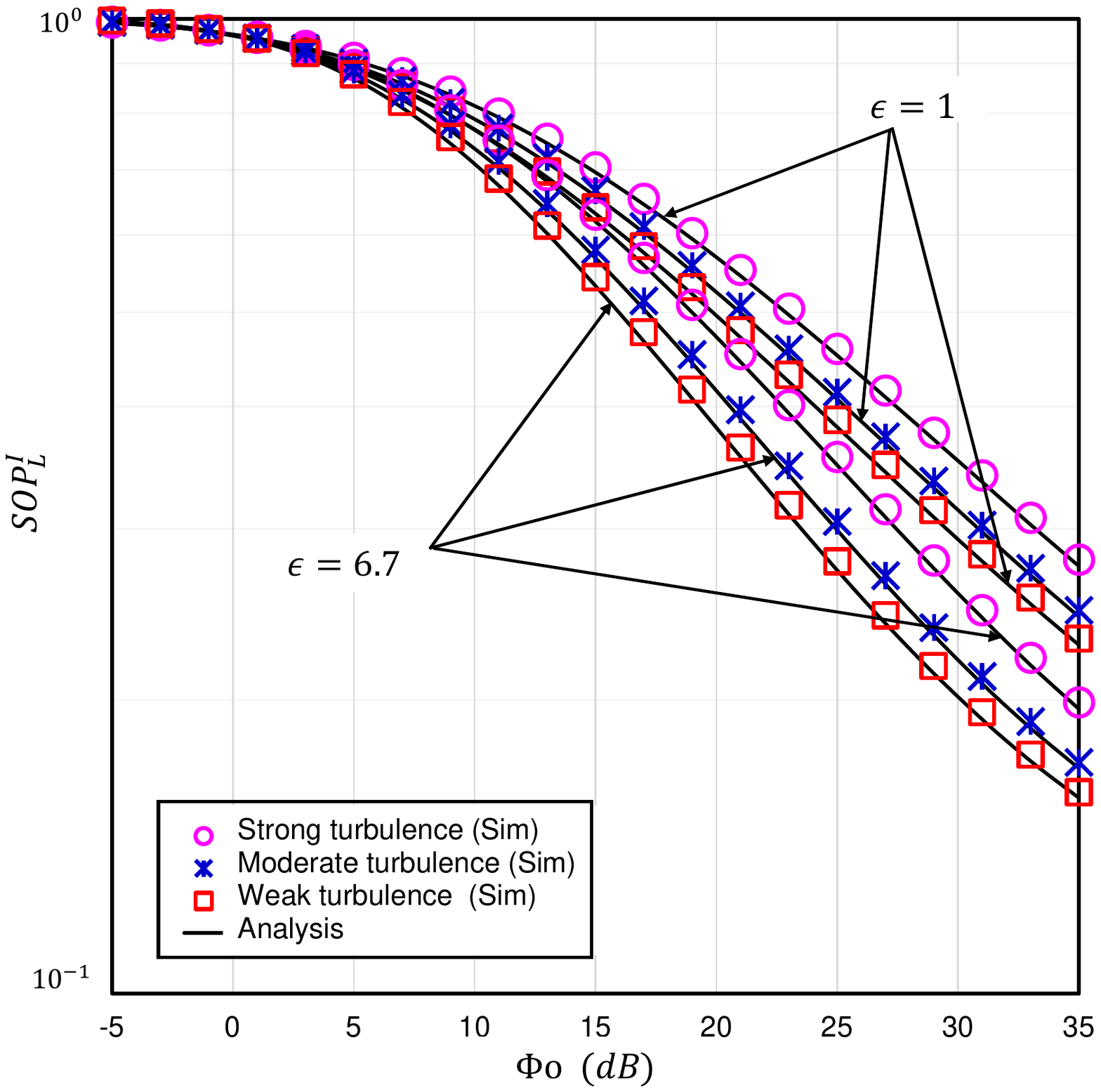}}
        \vspace{-10mm }
    \caption{
           The SOP versus $\Phi_{o}$ for selected values of $\alpha_{o}$, $\beta_{o}$ and $\epsilon$ with $\Omega_{o}=1$, $g=2$, $s=2$, $P_{o}=0.1$, $\alpha_{r}=\alpha_{p}=\alpha_{e}=2$, $\mu_{r}=\mu_{p}=\mu_{e}=6$,$\Phi_{r}=\Phi_{p}=15dB$, $\Phi_{e}=-5dB$, $\Psi_{Q}=-10dB$, and $\Upsilon_{e}=0.05$ bits/sec/Hz.
    }
    \label{g81}
\end{figure} 

Since stronger turbulence yields a weaker SNR at the destination, the weaker the turbulence, the better the secrecy performance that is also demonstrated in \cite{islam2020secrecy}. Moreover, a comparison between two detection techniques is also presented that clearly reveals that HD technique overcomes the atmospheric turbulence more appropriately than the IM/DD technique. The results in \cite{lei2020secure} also completely match with this outcome.

Besides the atmospheric turbulence, pointing error also plays a vital role on the FSO link performance that is illustrated in Fig. \ref{g81} by depicting SOP as a function of the average SNR of the FSO channel experiencing Scenario-I.

The result exhibits that higher pointing error (lower value of $\epsilon$) is more detrimental for better SOP than the lower pointing error (higher value of $\epsilon$). This happens since lower pointing error designates a better pointing accuracy that yields a better detection of the received signal. This statement can be easily justified with similar results presented in \cite{lei2020secure}.

\section{Conclusions}
\label{conclusions}

In this work, the secrecy performance of a CUN based hybrid RF / FSO network has been investigated wherein both maximum transmit power and interference power constraints within the CR spectrum are taken into consideration. Assuming all the RF and FSO links were subjected to $\alpha-\mu$ fading and M\'alaga ($\mathcal{M}$) turbulence, respectively, the expressions of SOP, probability of SPSC, and EST were derived in closed-form. The tightness of the simulation and analytical results were also demonstrated via MC simulations. Furthermore, the impacts of the system parameters on the security performance were analyzed capitalizing on the derived expressions. Numerical outcomes revealed that with decrease in fading severity of primary interference link ($S-P$ link) and wiretap link ($S-E$ link), and increase in link blockage probability, atmospheric turbulence and pointing error, the secrecy performance deteriorates significantly where the HD technique overcomes those secrecy threats more strongly relative to the IM/DD technique.

\appendices
\section*{Appendix}

\subsection{Proof of $\Lambda_{2}$}
\label{a1}

$\Lambda_{2}$ is given as
\begin{align}
\label{eqn:a2}
\nonumber
\Lambda_{2}&=\underbrace{\int_{\frac{\Psi_{Q}}{\Psi_{T}}}^{\infty}\frac{\alpha_{p}\delta_{p}^{\mu_{p}}}{2\Gamma(\mu_{p})}y^{\Theta_{p}}e^{-\delta_{p}y^{\tilde{\alpha_{p}}}}dy}_{\mathcal{P}_{1}}
\\
\nonumber
&-\underbrace{\int_{\frac{\Psi_{Q}}{\Psi_{T}}}^{\infty}\sum_{m_{r}=0}^{\mu_{r}-1}\frac{\alpha_{p}\delta_{p}^{\mu_{p}}}{2\Gamma(\mu_{p})}\frac{\delta_{r}^{m_{r}}}{m_{r}!}
\frac{\Psi_{Q}^{-\tilde{\alpha_{r}}m_{r}}\gamma_{r}^{\tilde{\alpha_{r}}m_{r}}y^{\Theta_{p}+\tilde{\alpha_{r}}m_{r}}}{e^{(\delta_{p}y^{\tilde{\alpha_{p}}}+\delta_{r}\Psi_{Q}^{-\tilde{\alpha_{r}}}\gamma_{r}^{\tilde{\alpha_{r}}}y^{\tilde{\alpha_{r}}})}}dy}_{\mathcal{P}_{2}}.
\end{align}
By using \cite[~Eq. (3.381.9)]{GR:07:Book}, $\mathcal{P}_{1}$ is expressed as
\begin{align}
\nonumber
\mathcal{P}_{1}&= \frac{\alpha_{p}\delta_{p}^{\mu_{p}}}{2\Gamma(\mu_{p})}\int_{\frac{\Psi_{Q}}{\Psi_{T}}}^{\infty}y^{\Theta_{p}}e^{-\delta_{p}y^{\tilde{\alpha_{p}}}} dy =\Xi_{5}^{II}.
\end{align}
$\mathcal{P}_{2}$ is written as
\begin{align}
\nonumber
\mathcal{P}_{2}&=\sum_{m_{r}=0}^{\mu_{r}-1}\frac{\alpha_{p}\delta_{p}^{\mu_{p}}\delta_{r}^{m_{r}}}{2\Gamma(\mu_{p})m_{r}!}
\Psi_{Q}^{-\tilde{\alpha_{r}}m_{r}}\gamma_{r}^{\tilde{\alpha_{r}}m_{r}}\int_{\frac{\Psi_{Q}}{\Psi_{T}}}^{\infty}y^{\Theta_{p}+\tilde{\alpha_{r}}m_{r}}
\\
&\times e^{-\delta_{p}y^{\tilde{\alpha_{p}}}}e^{-\delta_{r}\Psi_{Q}^{-\tilde{\alpha_{r}}}\gamma_{r}^{\tilde{\alpha_{r}}}y^{\tilde{\alpha_{r}}}}dy.
\end{align}
Due to tractable analysis, assuming $\tilde{\alpha_{p}}=\tilde{\alpha_{r}}$ and using similar formula as for $\mathcal{P}_{1}$, $\mathcal{P}_{2}$ is expressed as
\begin{align}
\label{eqn:p2-1}
\nonumber
\mathcal{P}_{2}&=\sum_{m_{r}=0}^{\mu_{r}-1}\frac{\alpha_{p}\delta_{p}^{\mu_{p}}\delta_{r}^{m_{r}}}{2\Gamma(\mu_{p})m_{r}!}
\Psi_{Q}^{-\tilde{\alpha_{r}}m_{r}}\gamma_{r}^{\tilde{\alpha_{r}}m_{r}}\int_{\frac{\Psi_{Q}}{\Psi_{T}}}^{\infty}y^{\Theta_{p}+\tilde{\alpha_{r}}m_{r}}
\\
\nonumber
&\times e^{-y^{\tilde{\alpha_{r}}}(\delta_{p}+\delta_{r}\Psi_{Q}^{-\tilde{\alpha_{r}}}\gamma_{r}^{\tilde{\alpha_{r}}})}dy=\sum_{m_{r}=0}^{\mu_{r}-1}\frac{\alpha_{p}\delta_{p}^{\mu_{p}}\delta_{r}^{m_{r}}}{2\Gamma(\mu_{p})m_{r}!}
\\
\nonumber
&\times \Psi_{Q}^{-\tilde{\alpha_{r}}m_{r}}\gamma_{r}^{\tilde{\alpha_{r}}m_{r}}\frac{\Gamma\left[\Omega,\left(\delta_{p}+\delta_{r}\Psi_{T}^{-\tilde{\alpha_{r}}}\gamma_{r}^{\tilde{\alpha_{r}}}\right)\left(\frac{\Psi_{Q}}{\Psi_{T}}\right)^{\tilde{\alpha_{r}}}\right]}{\tilde{\alpha_{r}}(\delta_{p}+\delta_{r}\Psi_{Q}^{-\tilde{\alpha_{r}}}\gamma_{r}^{\tilde{\alpha_{r}}})^{\Omega}}.
\end{align}
Utilizing the integral formula of \cite[~Eq. (8.352.7)]{GR:07:Book} and performing some mathematical manipulations, $\mathcal{P}_{2}$ is demonstrated in an alternative form as
\begin{align}
\nonumber
\mathcal{P}_{2}&=\sum_{m_{r}=0}^{\mu_{r}-1}\sum_{m_{3}=0}^{\Omega-1}\sum_{m_{4}=0}^{m_{3}}\sum_{m_{5}=0}^{\infty}\binom{m_{3}}{m_{4}}\binom{\Omega+m_{5}-1}{m_{5}}
\\
&\times \Xi_{6}^{II} \gamma_{r}^{\tilde{\alpha_{r}}(m_{r}+m_{4}+m_{5})}e^{-(\delta_{p}\Psi_{Q}^{\tilde{\alpha_{r}}}\Psi_{T}^{-\tilde{\alpha_{r}}}+\delta_{r}\Psi_{T}^{-2\tilde{\alpha_{r}}}\Psi_{Q}^{\tilde{\alpha_{r}}}\gamma_{r}^{\tilde{\alpha_{r}}})}.
\end{align}

\bibliographystyle{IEEEtran}
\bibliography{IEEEabrv,main.bib}

\end{document}